\documentclass[10pt,conference,compsocconf]{IEEEtran}
\IEEEoverridecommandlockouts
\usepackage{cite}
\usepackage{amsmath,amssymb,amsfonts}
\usepackage{graphicx}
\usepackage{textcomp}
\usepackage{xcolor}
\usepackage{subfig}
\usepackage{mathtools}
\usepackage{tikz}
\usepackage{mleftright}
\usepackage{nicefrac}
\usepackage{varwidth}
\usepackage{comment}
\usepackage{multirow}
\usepackage[noend]{algpseudocode}
\usepackage{algorithm}

\DeclarePairedDelimiter\ceil{\lceil}{\rceil}

\algnewcommand{\Pragmanoident}[1]{\State \(\textbf{\#pragma}\) #1}
\algnewcommand{\LineComment}[1]{\State \(\triangleright\) #1}
\algnewcommand\And{\textbf{and}}
\renewcommand*\Call[2]{\textproc{#1}(#2)}
\algnewcommand\algorithmicforeach{\textbf{for each}}
\algdef{S}[FOR]{ForEach}[1]{\algorithmicforeach\ #1\ \algorithmicdo}
\algnewcommand\algorithmicpragma{\textbf{\#pragma}}
\algdef{S}[FOR]{Pragma}[1]{\algorithmicpragma\ #1}

\AtBeginDocument{%
  \providecommand\BibTeX{{%
    \normalfont B\kern-0.5em{\scshape i\kern-0.25em b}\kern-0.8em\TeX}}}

\newcommand{\ralg}[1]{Algorithm~\ref{alg:#1}}
\newcommand{\rsec}[1]{Section~\ref{sec:#1}}
\newcommand{\rssec}[1]{Subsection~\ref{sec:#1}}

\newcommand{\rtab}[1]{Table~\ref{tab:#1}}
\newcommand{\rfig}[1]{Figure~\ref{fig:#1}}

\newcommand{\tit}[1]{{\textit{#1}}}

\newcommand*\circled[1]{\tikz[baseline=(char.base)]{
\node[shape=circle,draw,inner sep=1pt] (char) {#1};}}

\begin{document}

\title{Tensor Slicing and Optimization for Multicore NPUs}

\author{\IEEEauthorblockN{Rafael Sousa}
\IEEEauthorblockA{IC-UNICAMP \\
Campinas, Brazil \\
rafael.sousa@ic.unicamp.br}
\and
\IEEEauthorblockN{Marcio Pereira}
\IEEEauthorblockA{IC-UNICAMP \\
Campinas, Brazil \\
mpereira@ic.unicamp.br}
\and
\IEEEauthorblockN{Yongin Kwon and Taeho Kim}
\IEEEauthorblockA{Electronics and Telecommunications \\ Research Institute (ETRI) \\
Seoul, Republic of Korea \\
\{yongin.kwon, taehokim\}@etri.re.kr}
\and
\IEEEauthorblockN{Namsoon Jung and Chang Soo Kim}
\IEEEauthorblockA{SilicoNeuro/AiM Future \\
Seoul, Republic of Korea \\
\{yongin.kwon, changsoo.kim\}@siliconeuro.com}
\and
\IEEEauthorblockN{Michael Frank}
\IEEEauthorblockA{MagiCore \\
Santa Clara, USA \\
michael@magicore.com}
\and
\IEEEauthorblockN{Guido Araujo}
\IEEEauthorblockA{IC-UNICAMP \\
Campinas, Brazil \\
guido@unicamp.br}
}

\maketitle

\begin{abstract}
Although code generation for Convolution Neural Network (CNN) models has been extensively studied, performing efficient data slicing and parallelization for highly-constrai\-ned Multicore Neural Processor Units (NPUs) is still a challenging problem. Given the size of convolutions' input/output tensors and the small footprint of NPU on-chip memories, minimizing memory transactions while maximizing parallelism and MAC utilization are central to any effective solution. This paper proposes a TensorFlow XLA/LLVM compiler optimization pass for Multicore NPUs, called Tensor Slicing Optimization (TSO), which: (a) maximizes convolution parallelism and memory usage across NPU cores; and (b) reduces data transfers between host and NPU on-chip memories by using DRAM memory burst time estimates to guide tensor slicing. To evaluate the proposed approach, a set of experiments was performed using the NeuroMorphic Processor (NMP), a multicore NPU containing 32 RISC-V cores extended with novel CNN instructions. Experimental results show that TSO is capable of identifying the best tensor slicing that minimizes execution time for a set of CNN models. Speed-ups of up to 21.7\% result when comparing the TSO burst-based technique to a no-burst data slicing approach. To validate the generality  of the TSO approach, the algorithm was also ported to the Glow Machine Learning framework. The performance of the models were measured on both Glow and TensorFlow XLA/LLVM compilers, revealing similar results.
\end{abstract}


\vspace{-0.2cm}

\section{Introduction}
\label{sec:introduction}

Deep Learning using Convolutional Neural Network (CNN) has become a significant architecture model technique that considerably increases the accuracy on many modern AI applications. The steady increase in the adoption of CNNs is driven mostly by applications in the Computer Vision domain where it addresses problems like Object Recognition \cite{simonyan2014very,he2016deep,zhou2014learning}, Object Detection \cite{girshick2014rich,sermanet2013overfeat}, and Video Classification \cite{karpathy2014large,sermanet2013pedestrian}. Other areas, like Speech Recognition and Natural Language Processing (NLP) have also benefited from the application of CNN models \cite{abdel2014convolutional,kim2014convolutional}. 

Followed by its accuracy improvements, the size and complexity of state-of-the-art CNNs have also grown significantly. For instance, LeNet-5 \cite{lecun1998gradient}, a model that recognizes handwritten digits, has less than 1 Million parameters, while more complex models, like InceptionV3 \cite{szegedy2016rethinking} which classifies thousands of different object categories, has more than 23 million parameters. Such increase in the model complexity and parameters size, not only demands more computational power but also produces a  significant increase in the data movement between host (off-chip) and the AI accelerator (on-chip) memories thus considerably impacting energy-consumption \cite{tu2018rana} and memory traffic.

It is well-known that convolution is the most expensive operation of a CNN,  accounting for the largest share of a CNN execution. Given the size of its tensor inputs and the wide variety of configuration parameters (e.g., kernel size, stride, etc), selecting the best data mapping which maximizes convolution parallelism while minimizing memory transactions is a key factor to the performance of any AI accelerator. This is particularly critical for multicore Neural Processing Units (NPUs), which have stringent (on-chip) memory constraints and need to achieve large inference throughput. 

\begin{figure}[t]
  \centering
  \includegraphics[scale=0.46]{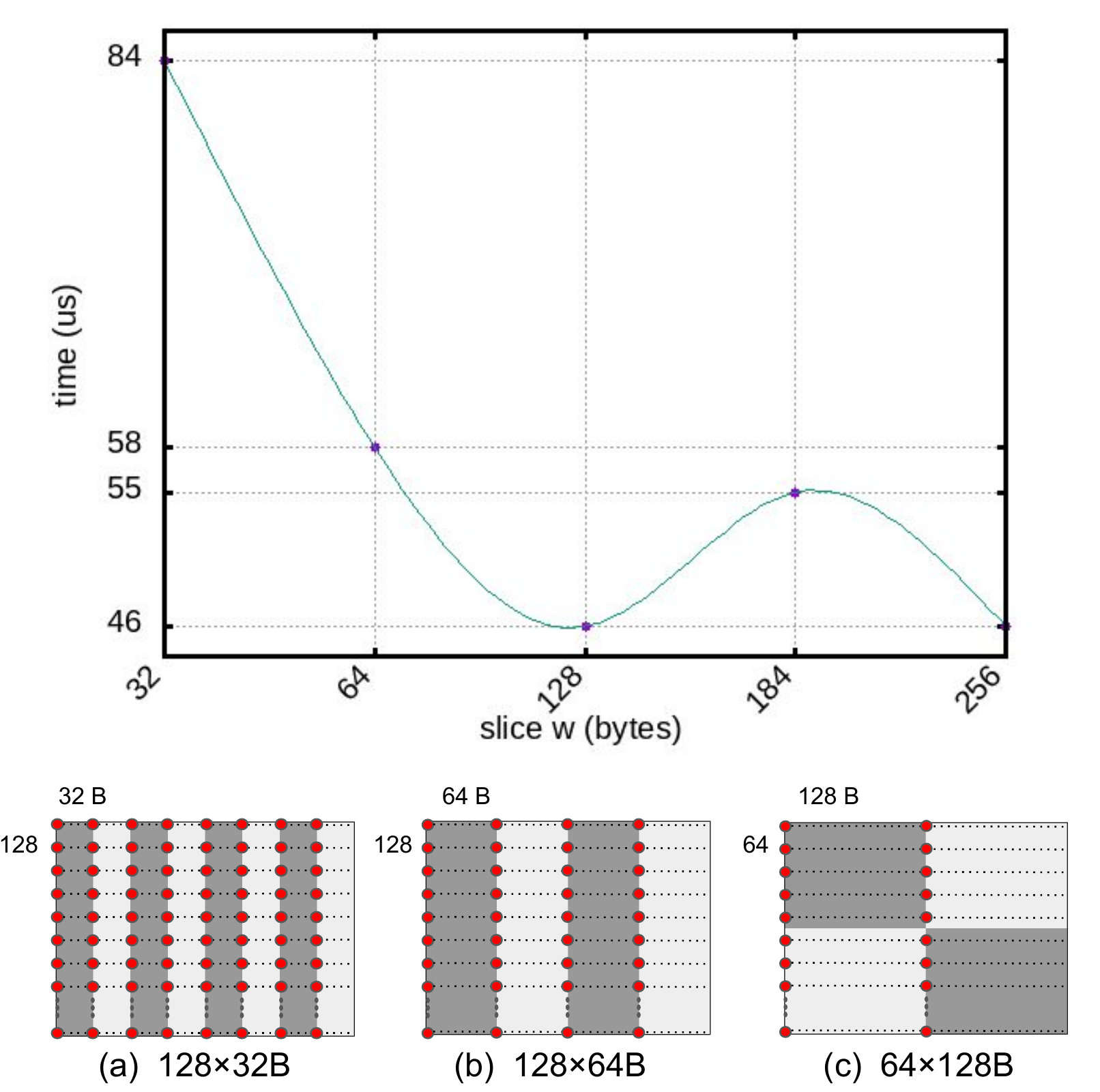}
  \caption{Memory access with different slice shapes.}
  \label{fig:tile_shapes_intro}
\end{figure}

To achieve that, Convolution input tensors and weights need to be divided into slices that fit into NPU on-chip memories. Slices are brought from (slow) external DRAM to (fast) on-chip memories. Input tensors and weight slices are then used to perform Convolution, one set of slices at a time. Depending on how the slice shapes and sizes are selected, the convolution execution time can drastically change. As an example, consider  Figure \ref{fig:tile_shapes_intro}, which shows the time taken by a Convolution when using slices of different shapes. In that example, the input tensor is a single channel with $128 \times 128$ 16-bit fixed-point elements (row-major) computed over a single kernel of size $1 \times 1$. In the figure, slices are represented as light/dark gray areas, and each red dot represents a (128B) \textit{memory burst} access to the DRAM. Accessing time in a  DRAM can be divided into two components: (a) CAS latency, which is the time taken to read the first byte of a memory burst from the DRAM Row Buffer; and (b) Access latency, which is the time taken to read the following bytes of the burst. For example, reading the first byte from a 128B  burst of a typical DDR3 memory takes $\sim$ 14ns, the same time it takes to read all the remaining 127 bytes of that burst. Depending on how data is sliced, memory bursts can have an enormous impact on execution time. For example, in Figure \ref{fig:tile_shapes_intro} the Convolution can be divided into: (a) 8 $128\times32B$ slices resulting in 1024 bursts (red dots) and an execution time of 84us; (b) 4  $128\times64B$ slices corresponding to 512 bursts and a reduced 58us execution time; and (c) 4 $64\times128B$ slices which require 256 bursts and 46us execution time, a 45\% reduction in the convolution time when comparing to the slicing in (a). In Figure \ref{fig:tile_shapes_intro}, slicing (c) is represented by the smallest memory access time at w = 128B.  From that point on, as the width (w) of the slice continues to increase, memory access time worsens and then improves again at the next memory burst alignment (w = 256).

Although memory access coalescing is a common problem in GPU code generation, it has not been explored in the context of multicore NPU parallelism. This paper proposes a compiler optimization for multicore NPUs, called Tensor Slicing Optimization (TSO), which has two goals: (a) to maximize the parallelization of convolutions across the memories of the available NPU cores; and (b) to reduce data transfers between host and the cores' on-chip memory. This is achieved by modeling, at compile time, the memory utilization of the various NPU cores in the search for the best input/output tensor slicing which minimizes data transfers between the host and the NPU cores' memories. To evaluate this approach, a set of experiments was performed using the NeuroMorphic Processor (NMP), a multicore NPU containing 32 cores, and the TensorFlow XLA LLVM compiling toolchain.

This paper is divided as follows. \rsec{background} provides a background review. \rsec{nmp} describes details about the NMP accelerator. \rsec{mapping_strategies} shows how to map Convolution Layers on NMP using the TSO algorithm. \rsec{tf_xla} describes the compilation flow using the TF-XLA compiler. \rsec{results} shows the experimental results. \rsec{related_works} analyzes the works related to this paper, and finally, \rsec{conclusion} concludes the work.

\section{Background} 
\label{sec:background}

A CNN model can be seen as a directed acyclic graph composed of multiple layers of operations in which a set of input channels (e.g., images) is processed. After been processed through a very deep hierarchy of layers, an output results. The output result is usually an array composed of a probabilistic distribution that  classifies  the  given  image  into  a  set  of  classes  (e.g.,  dog,  cars,  etc).  Our focus on this work is on inferences applied to already pre-trained models.

Among  all  possible  layers  that  compose  a  CNN,  the  Conv-layer  account for  more  than  90\%  of  the  execution  time  \cite{cong2014minimizing}  of  a  model,  and  generates  a large amount of data movements. This is especially critical on architectures with small on-chip memories like NPUs that are used to perform inference on mobile and embedded devices. Data tiling approaches like the one proposed herein are thus a key technique to reduce memory transactions on such devices.

\begin{figure} [t]
  \centering
  \includegraphics[scale=0.26]{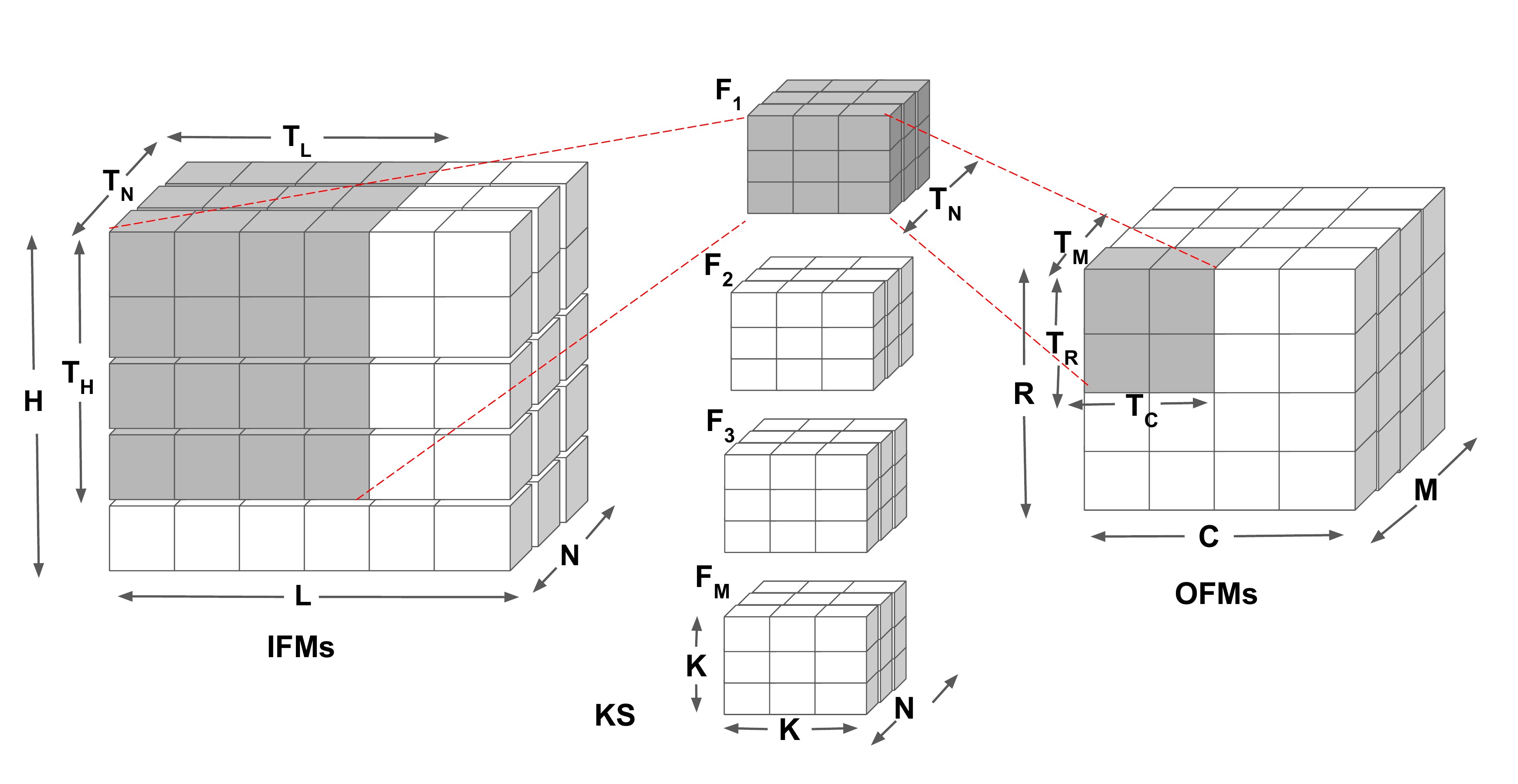}
  \caption{A Tiled Convolutional Layer.}
  \label{fig:cnn_layers}
\end{figure}

As shown in \rfig{cnn_layers}, a Conv-layer uses $N$ input feature maps (\textit{IFM}) of size $H$ (height) $\times$ $L$ (width) and a set of  pre-trained weights. The weight set ($KS$) is a set of $M$ multidimensional (e.g., 3-D) arrays kernels/filters of size $N \times K \times K $. Each filter slides over the IFMs performing a 3-D convolution with a stride factor of $S$. After sliding over the entire input image (IFMs), an $R \times C$ output feature map (\textit{OFM}) is generated. A set of $M$ OFMs results after applying all $M$ filters in $KS$ to the IFMs.

Different Conv-layers in a CNN usually have different kernel sizes, different numbers of IFM and OFM with distinct sizes, and variable strides. For instance, the first Conv-layer from Inception-V3 \cite{szegedy2016rethinking} has $3\times299\times299$ elements in its input. If we consider the data type of the input as 16-bit fixed-point, the size of the input becomes 524KB. This is impractical to store at once into the constrained on-chip memories available on typical NPUs. Because of that, data tiling is a  mandatory task in the computation of a Conv-layer.

Data tiling of a Conv-layer consists in partitioning its IFM and OFM data maps into small tiles and dividing the filters in $KS$ so that each $IN$, $OUT$ and $W$ tiles fit together at the same time into the NPU on-chip memories. Figure \ref{fig:cnn_layers} shows how the IFM and OFM data maps are respectively divided into $IN \ = (T_N, T_H, T_L $) tiles,  and  $OUT \ = (T_M, T_R, T_C)$ tiles. Notice that the dimensions $T_H$ and $T_L$ of the IFM can thus be computed using the dimensions $T_R$ and $T_C$ of the OFM through Equation \ref{eq:infromout}.
\begin{equation}
\begin{split}
    T_H = (T_R - 1)S + K \\
    T_L = (T_C - 1)S + K
\label{eq:infromout}
\end{split}
\end{equation} 
where $K$ and $S$ are the kernel size and stride, respectively. Moreover, if the filters/kernels in $KS$ do not fit into their respective NPU on-chip memories, $KS$ is also partitioned into $W \ = (T_M, T_N, K, K)$ tiles, where $T_M$ is the number of filters and $T_N$ the number of channels to be loaded from each filter.  

Different tile shapes can be explored when partitioning a convolution to execute on an NPU. Each tile shape leads to different memory accesses and usage of the resources available on the NPU. Besides that, different scheduling strategies can be explored, each one with a specific memory access pattern that leads to different data-reuse. The way computation is mapped can considerably affect the data movement between host and NPU memories leading to a poor data re-use. Moreover, if not properly done, tiling can also result in a poor utilization of the NPU’s Multiplier Accumulator (MAC) units, which can become idle during the computation  (more details in Section \ref{sec:mapping_strategies}).

\section{The NMP Architecture}
\label{sec:nmp}

Although CPUs have been proposed to accelerate CNNs by relying on multicore parallelism and SIMD instructions \cite{vanhoucke2011improving,lee2018efficient}, the number and complexity of the layers in modern CNN models make it very difficult to run the entire network on CPUs. To improve inference throughput, (fast) GPU solutions have been proposed to process a large amount of data \cite{chetlur2014cudnn,song2016bridging}. Field Programmable Gate Arrays (FPGAs), on the other hand, have been extensively used as an alternative to this problem as they offer good performance and reconfigurability \cite{sankaradas2009massively,peemen2013memory,chen2016eyeriss,chakradhar2010dynamically,gokhale2017snowflake}. Nevertheless, these architectures are not efficient power-performance solutions for critical edge applications, like surveillance cameras and cellphone face recognition, etc., which have stringent execution and power consumption constraints.  Several types of accelerators have been proposed to accelerate CNNs in a power-efficient way. Specialized ASICs \cite{cavigelli2015origami}, Neural Processing Units (NPUs) \cite{liu2015reno,kim2016neurocube}, and Tensor Processing Units (TPUs) \cite{jouppi2017datacenter} are some examples. 

This paper uses the  NeuroMorphic Processor (NMP) by  LG Electronics (LGE) as a compiling target.  The key idea behind the NMP  architecture is to use RISC-V ISA Extensions to design relevant CNN instructions like Conv-layers, FC-layers, Pooling layers, Element Wise operations, etc.

\begin{figure}[t]
  \centering
  \includegraphics[scale=0.35]{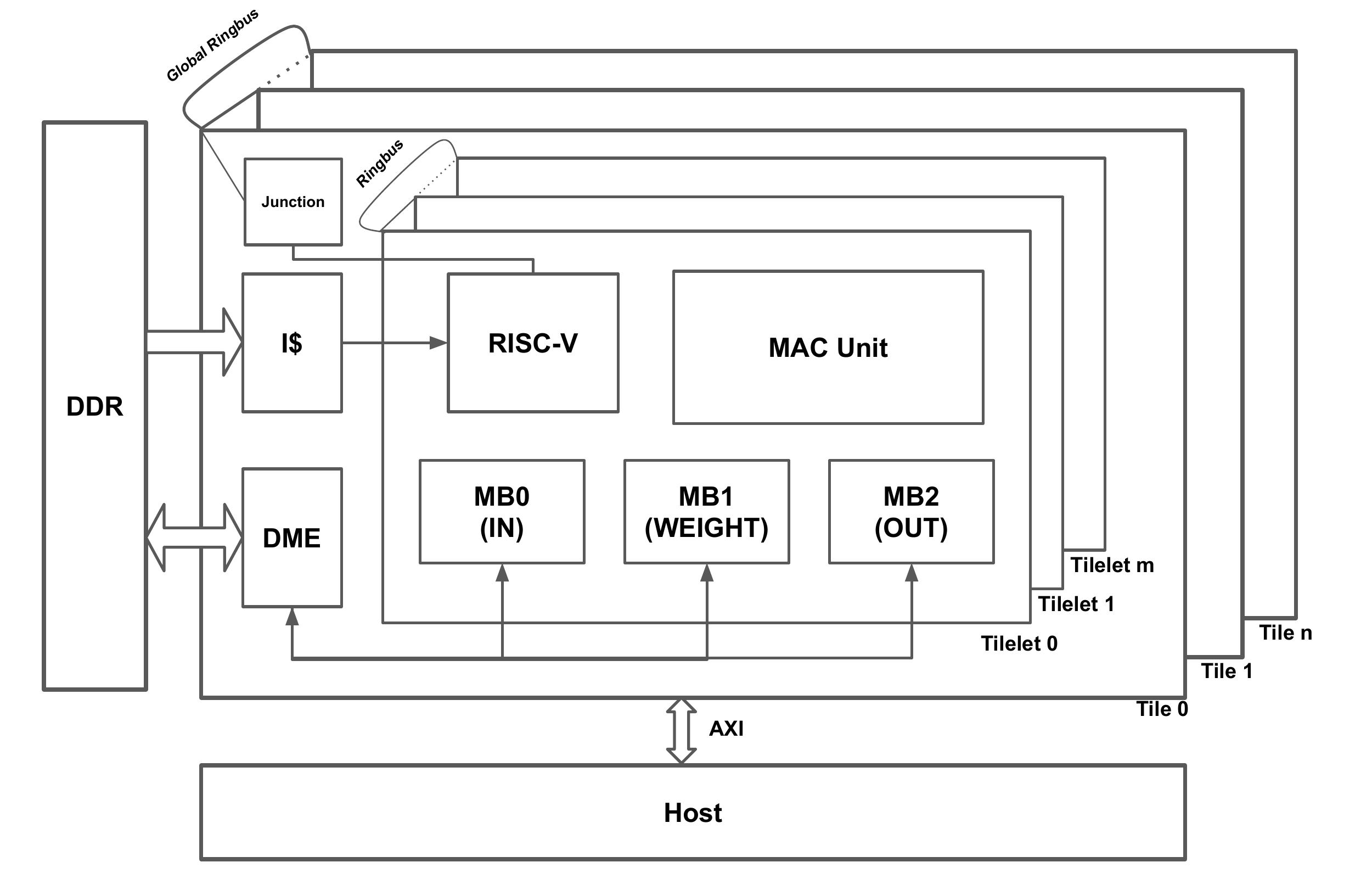}
  \caption{NMP Architecture.}
  \label{fig:nmp_arch}
\end{figure}

The NMP architecture (\rfig{nmp_arch}) is a multicore NPU that contains an ARM57 processor that works as a host for a set of multiple Tile (TLE) processors, containing each a set of Tilelet (TLT) cores. Each TLT  has one RISC-V core, three on-chip (scratchpad) memories, namely MB0, MB1 and MB2, which respectively store the $IN$, $W$ and $OUT$ tiles from the IFM, KS and OFM data maps. Besides that, each TLT is also equipped with a MAC acceleration unit to execute CNN operations. The MAC unit execution is triggered by the RISC-V core and is capable of executing 8- and 16-bit fixed-point operations with the memory layout organized in NCHW format. The datapath between the MAC unit and TLT on-chip memories (MBLOBs - MB0, MB1 and MB2) is 128-bit wide which means that for 8-bit fixed-point,  up to 16 MAC operations are executed per cycle, while for 16-bit fixed-point, up to 8 MAC operations per cycle may be executed.

The data transfers between NMP and host happens through a Data Movement Engine (DME) module, as shown in Figure \ref{fig:nmp_arch}. The host communicates with the NMP through an AXI interface and data can be shared between TLTs of different TLEs by using a Global Ringbus. The TLTs of a TLE also have their own Ringbus to communicate data between them. The instructions executed by the RISC-V cores are fetched from the host memory and stored into a cache instruction, which is shared between the TLTs of the same TLE.

To execute any computation on NMP, the model is first compiled using the TF-XLA compiler. The execution of the compiled model starts on the RISC-V cores of each TLT, each one of them executing independently of each other. To invoke any computation on the MAC units, one of the following extended instructions are run by the RISCV-core: nmp\_conv2d, nmp\_veop, nmp\_pool, nmp\_acti\-va\-tion and nmp\_percept. To load data from the DRAM to the TLT on-chip memory, the RISC-V has to execute a data movement instructions (nmp\_load and nmp\_load3d) so that the DME is invoked to do the job. Data is brought back from each TLT on-chip memory to the host DRAM by executing the RISC-V extended nmp\_store and nmp\_store3d instructions.

Despite the fact that each one of the TLTs of a single TLE executes its computation independently of the others, NMP has support to a special \tit{multicast} load instruction which executes on a single TLT but enables all TLTs of that TLE to load the same data from the DRAM. As an example, suppose that all TLTs of a given TLE process the same IFMs but using different filters each. Instead of having to load the (IFMs) IN tiles multiple times, one for each TLT of the TLE, a multicast load can be used to load all TLTs with the same IN tile, thus allowing it to be processed by different filters in parallel, for example. Notice that TLTs from distinct TLEs cannot use the same multicast to load the same data. In such case, multiple multicast loads are required, one for each TLE, to load the same data to all its TLTs.

Three levels of hardware-based semaphores are available in NMP to perform synchronization among TLEs and TLTs. They enable the following operations: (a) synchronize computation inside a TLT (e.g., after invoking the MAC unit to execute an operation, it is possible to block the RISC-V execution until the unit finishes its work); (b) synchronize computation between TLTs of the same TLE (e.g.,  during a multicast load, all TLTs of a specific TLE are blocked until their corresponding on-chip memories receive the data from the DRAM); and (c) synchronize computation between TLTs of different TLEs (e.g., assume for a given model that a layer has a dependency on its predecessor layer; if a TLT finishes its computation before all other TLTs working on the same layer, it must be blocked so it can not proceed to the following layer).

The NMP architecture used in this work is composed of 4 TLEs, each containing 8 TLTs (RISC-V + MAC unit). Each TLT has three on-chip memories of size 8KB each. With an operating frequency of @1GHz, NMP (all the 32TLTs together) has a theoretical performance peak of either 512- or 256-GMACs/sec when executing 8- or 16-bit fixed-point, respectively. The DRAM memory is a DDR3 that operates on a 1066MHz clock rate (DDR3-2133 -- 17GB/s). For this edition of the architecture, NMP does not enable MAC/LOAD overlap. For future NMP-architectures, we anticipate the addition of extra on-chip (dual-port) memories that will allow the compiler to software-pipeline MAC/LOADs.

\section{NMP Mapping Strategies}
\label{sec:mapping_strategies}

There are different ways to transfer data from the host to an accelerator memory during the execution of a Conv-layer. If not done properly, the number of data movements may considerably increase, thus impacting performance and energy consumption \cite{tu2018rana}.  This paper proposes a search space exploration optimization algorithm called \textit{Tensor Slicing Optimization} (TSO) that seeks to identify the best TLE/TLT  data partitioning that minimizes the number of memory transfers during the execution of Conv-layers. 

\subsection{TSO Algorithm}

TSO works by  exhaustively exploring the solution space in the search for the best convolution tiling/scheduling strategy that minimizes execution time.  It first slices the input tensor of the convolution (IFMs) and its corresponding filters (KS) among the TLE processors of the NMP so that each TLE  computes a different slice of the Conv-layer's output. After that, each TLE slice is further partitioned into multiple tiles so as to distribute the computation among the TLT cores of the corresponding TLE processor. These two steps of the TSO algorithm are detailed below. Before moving further please consider from now on that every mention to \textit{slice} refers to a TLE data partitioning and every mention to \textit{tile} refers to a TLT partitioning.

\begin{algorithm}[!t]
\small
\begin{algorithmic}[1]
\caption{Select best TLE/TLT mapping}
\Function{TSO}{CONVS, \#TLE, \#TLT}
\Pragmanoident omp parallel
\Pragmanoident omp single
\ForEach {$conv \in CONVS$}
    \Pragma omp task
    \LineComment Let $conv = (IFM,KS,OFM)$
    \State $map[conv].bestTile.time \gets \infty$
    \LineComment Let $PART_{TLE} = \{KS,KS\&OFM,OFM\}$
    \ForEach{$p \in PART_{TLE}$}
        \LineComment Let $slice = (TLE_{R},TLE_{W})$
        \State $slice \gets$ $\Call{TLESlicing}{p, conv, \#TLE}$
        \LineComment Let $PART_{TLT} = \{IS,OS,WS\}$
        \ForEach{$q \in PART_{TLT}$}
            \LineComment Let $tile = (IN,W,OUT,time,schedule)$
            \State $tile \gets$ $\Call{TLTTiling}{q, conv, slice, \#TLT}$
            \If{$tile.time < map[conv].bestTile.time$}
                \State $map[conv].bestSlice \gets slice$
                \State $map[conv].bestTile \gets tile$
            \EndIf
        \EndFor
    \EndFor
    \EndFor
\EndFor
\State \Return $map$
\EndFunction
\label{alg:sel_conv_mapping}
\end{algorithmic}
\end{algorithm}

Initially (refer to \ralg{sel_conv_mapping}), TSO takes as input the set of convolutions of the model ($CONVS$)  and the number of TLEs ($\#TLE$) and TLTs ($\#TLT$) of the architecture (line 1). It then iterates over all convolutions (line 4) and initializes a $map$ which stores the best TLE slice and TLT tile for that specific convolution (lines 7). Then for all possible TLE slicing strategies $p$ available in $PART_{TLE}$ (line 9, see \rssec{TLEPart} for details), the algorithm uses a call to function $TLESlicing$ to divide the convolution IFMs and KS data across the TLEs. $TLESlicing$ returns tuple $slice = (TLE_{R}, TLE_{W})$, where $TLE_{R}$ refers to the part of the OFM (rows) that is generated from the slice of the IFMs designated to the TLE processor, and $TLE_{W}$ a subset of the KS filters that will run on that TLE processor.

Remember that each TLE processor in NMP has a set of TLT cores, and thus for each TLE slice produced in line 11, the slice data needs to be divided among its corresponding TLT cores. Hence, for each TLT scheduling strategy $q$ (line 13, see \rssec{TLTPart} for details), TSO computes the best TLT tile for the current TLE data slice using a call to $TLTTiling$ (line 15). This function takes as input the TLT scheduling strategy $q$, the convolution data ($conv$), the current TLE $slice$, and the number of TLTs ($\#TLT$). It then determines the best tiling of the TLE data among the TLT cores. The $TLTTiling$ function returns tuple $tile = (IN,W,OUT,time,schedule)$, where $IN$, $OUT$ are the tiles of the $IFM$ and $OFM$ data maps assigned to the TLTs of that TLE, and $W$ is a tile that contains a subset of the filter in $KS$.

The tuple also returns an estimate of the time taken to compute the convolution using that specific combination of TLE slice and TLT tile for the best possible scheduling ($schedule$) strategy (see \rssec{burst_modeling} for details). To achieve that, it takes into consideration the cost to load the $IN$ and  $W$ tiles from DRAM into the (on-chip) TLT memories  MB0 (IN) and MB1 (W), respectively, and the time to store the $OUT$ from the MB2 TLT (OUT) memory back to the host DRAM. Moreover, $time$  also includes the time taken by each evaluated partitioning to run on the MAC Unit using the various scheduling alternatives (see \rssec{Scheduling} for details).

After returning from $TLTTiling$, TSO compares (line 16) the estimated $time$ for the evaluated partitioning with the best time ($map[conv].bestTile.time$) found so far for that specific convolution. It then  stores it into the appropriate  $map$ entry (i.e., $map[conv]$) the corresponding TLE slice (line 17) and TLT tile (line 18). Finally, the $map$ containing the best slices/tiles for each convolution is then returned (line 19), so it can be used later by the code generator to synthesize and schedule the code for the TLT cores.

TSO is an algorithm that exhaustively explores the  solution space of all possible tensor slicing solutions for each Convolution of a model. As such,  it may take a long time to be executed, particularly when the CNN has a large number of Convolutions. Given that estimating the execution time of a  Convolution is independent of the others, the process of exploring the solution space is highly parallel. In this work, we  use OpenMP task-parallelism to accelerate this exploration, by running the simulation of the execution time of all Convolutions in parallel.  The parallel execution starts at line 2 with the creation of a thread pool. At this point of the execution, a unique thread is selected from the thread pool (line 3) to create a task for each Convolution (line 5). The tasks are then distributed across the threads within the thread pool to compute the TLE/TLT data partitioning and scheduling of the  Convolutions in parallel. 

\subsection{TLE Partitioning}
\label{sec:TLEPart}

The first step in the TSO optimization is to divide the filters in KS among the TLEs and define which part of the OFM (rows) the selected filters will compute. This is done according to the partitioning set defined in  \ralg{sel_conv_mapping} -- $PART_{TLE} = \{KS,KS\&OFM, OFM\}$, where KS, KS$\&$OFM and OFM are partitioning strategies computed  by \ralg{tleslice}.

\begin{algorithm}[!t]
\small
\begin{algorithmic}[1]
\caption{TLE Slicing}
\Function{TLESlicing}{p, conv, \#TLE}
    \State $rows \gets conv.R$
    \State $filters \gets conv.M$
    \If{$p = KS$}
        \State $TLE_W = \ceil{filters/\#TLE}$
        \State $TLE_R = rows$    
    \EndIf
    \If{$p = KS\&OFM$}
        \State $TLE_W = \ceil{filters/(\#TLE/2)}$
        \State $TLE_R = \ceil{rows/(\#TLE/2)}$ 
    \EndIf
    \If{$p = OFM$}
        \State $TLE_W = filters$
        \State $TLE_R = \ceil{rows/\#TLE}$
   \EndIf
\State \Return{$(TLE_R,TLE_W)$}
\EndFunction
\label{alg:tleslice}
\end{algorithmic}
\end{algorithm}

\textbf{KS partitioning} -- In the first partitioning scheme (line 4), only the convolution filters in KS are divided into slices among the TLEs (line 5). In terms of data replication, all the $R \times C$ elements of an OFM have to be computed by the TLE, which requires loading the entire IFMs at runtime on each TLE. This partitioning scheme usually works well on the last Conv-layers of a CNN model, given the increase in the number of filters as well as in their channels' depth. Thus, dividing the filters may reduce data transfers between DRAM and NMP.

\textbf{KS and OFM partitioning} -- The second partitioning scheme (line 7) divides both the filters and the OFM rows among the TLEs. For the NMP used in this work  (\#TLE = 4), it slices the OFM rows into two sets as well as the filters in KS, which are then combined to generate one slice for each TLE. This TLE partitioning scheme reduces the data transfer over the IFMs, compared to the first TLE partitioning scheme, but increases the loads over the filters, given that more filters are assigned to the slices. This scheme usually works better when both the KS and IFMs data have similar sizes.

\textbf{OFM partitioning} --  Finally, the third partitioning only divides the OFM rows among the TLEs (line 12). Given that the filters in KS have to be loaded by each TLE, this partitioning scheme usually works better on the first Conv-layers, since the IFMs are bigger when  compared to the filters in KS.

Since the NMP board used to collect the experiments for this paper does not have a global shared-buffer (shared among the TLEs), we have not considered this feature in designing TSO. However, TSO can be easily extended to consider a global shared-buffer since different slices of the IFMs/KS from different TLEs may be the same.

\subsection{TLT Partitioning}
\label{sec:TLTPart}

After choosing a TLE partitioning scheme, the workload of each TLE is divided among their corresponding TLTs by means of a call to function $TLTTiling$ in line 12 of \ralg{sel_conv_mapping}. $TLTTiling$ takes as input the TLT scheduling strategy ($q$), the convolution data ($conv$), the TLE slice ($slice$) resulting in line 8 of \ralg{sel_conv_mapping} and the number of TLTs at each TLE ($\#TLT$). It then produces as output the tuple $(IN,W,OUT,time,schedule)$ which will be used to generate code for the TLTs.

\begin{algorithm}[!t]
\small
\begin{algorithmic}[1]
\caption{TLT Tiling}
\Function{TLTTIling}{q,conv,slice,\#TLT}
    \LineComment  Let $tile = (IN,W,OUT,time,schedule)$
    \State $bestTile.time \gets \infty$
    \For{$T_R \gets 1$ to $slice.TLE_{R}$}
        \For{$T_C \gets 1$ to $conv.C$}
            \For{$T_N \gets 1$ to $conv.N$}
                \State $T_M \gets$ \Call{GetFilters}{$T_R, T_C, q, slice.TLE_W,\#TLT$}
                \State $W \gets$ \Call{GenTILE$_W$}{$T_M, T_N, conv, q$}
                \State $OUT \gets$ \Call{GenTILE$_{OUT}$}{$T_M,T_R, T_C$}  
                \State $IN \gets$ \Call{GenTILE$_{IN}$}{$T_N,T_R,T_C,conv$}
                \State $(time, schedule) \gets$ \Call{CalcTime}{$IN,W,OUT,q$}
                \If{$time < bestTile.time$}
                    \State $bestTile \gets (IN, W, OUT, time, schedule)$
                \EndIf    
            \EndFor
        \EndFor
    \EndFor
    \State \Return{$bestTile$}
\EndFunction
\label{alg:tlttile}
\end{algorithmic}
\end{algorithm}

Initially (refer to \ralg{tlttile}), $TLTTiling$ initializes variable $bestTile.time$ with infinity as it will store the smallest (estimated) execution time of all possible tiles visited by the function. To achieve that, a sequence of three nested loops (lines 4-6) generate the values $T_R$, $T_C$ and $T_N$ that are used to explore all possible IN, W and OUT tiles shapes that can be formed from a TLE slice. But before computing the IN and OUT tiles for that TLE slice, the convolution filters in $TLE_W$ need to be divided among the various TLTs. This is done in line 7, which also determines the maximum number of filters ($T_M$) that can fit into the MB1 ($W$) memory of a TLT, and in line 8, which generates the corresponding W tile. In the case of an unbalanced filter partitioning, the remaining filters are spread among the TLTs which have the lowest IDs. This is followed by calling functions to generate the $OUT$ tile ($GenTILE_{OUT}$ in line 9) and $IN$ tile ($GenTILE_{IN}$ in line 10). These two functions also check if the tiles $OUT$ and $IN$ respectively fit into memories MB2 (OUT) and MB0 (IN) of a TLT, as shown in Equation \ref{eq:tilesizes}, where type stands for either 8- or 16-bit fixed-point. The functions between lines 8-10 also calculate the number of times the IN, OUT and W tiles have to loaded/stored from/to the DRAM to cover all the workload of a TLE slice (more details in Subsection \ref{sec:TLTPart}).
\begin{equation}
\centering
\begin{cases}
  IN = T_N \times T_H \times T_L \times type \leq MB0 \\
  W = T_M \times T_N \times K \times K \times type \leq MB1 \\
  OUT = T_M \times T_R \times T_C \times type \leq MB2 
\label{eq:tilesizes}
\end{cases}
\end{equation}

Tiles $IN$, $W$, $OUT$ and the tiling strategy $q$, are then passed to function $CalcTime$ (line 11), so it can estimate the best $schedule$ and $time$ to compute the TLE slices using the generated tiles (more details in Subsection \ref{sec:burst_modeling}).  Finally, the algorithm tests if the $time$ computed for the current tiling is smaller than the $bestTile.time$ seen so far, and if so, it updates the $bestTile$.

\subsection{Scheduling}
\label{sec:Scheduling}

With the data divided among the TLEs/TLTs, different scheduling strategies -- Input Stationary (IS), Output Stationary (OS) and Weight Stationary (WS) --  may be used by the TLT cores to execute the convolution, each providing a different memory access pattern ($schedule$). Given that the data-transfers of the mapping strategies presented herein can be determined statically, we compute the number of accesses to the DRAM required by each one of them, according to their data-flow patterns, so as to determine the one that can result in the best data reuse.

\begin{figure}[ht]
  \centering
  \fbox{\includegraphics[scale=0.25]{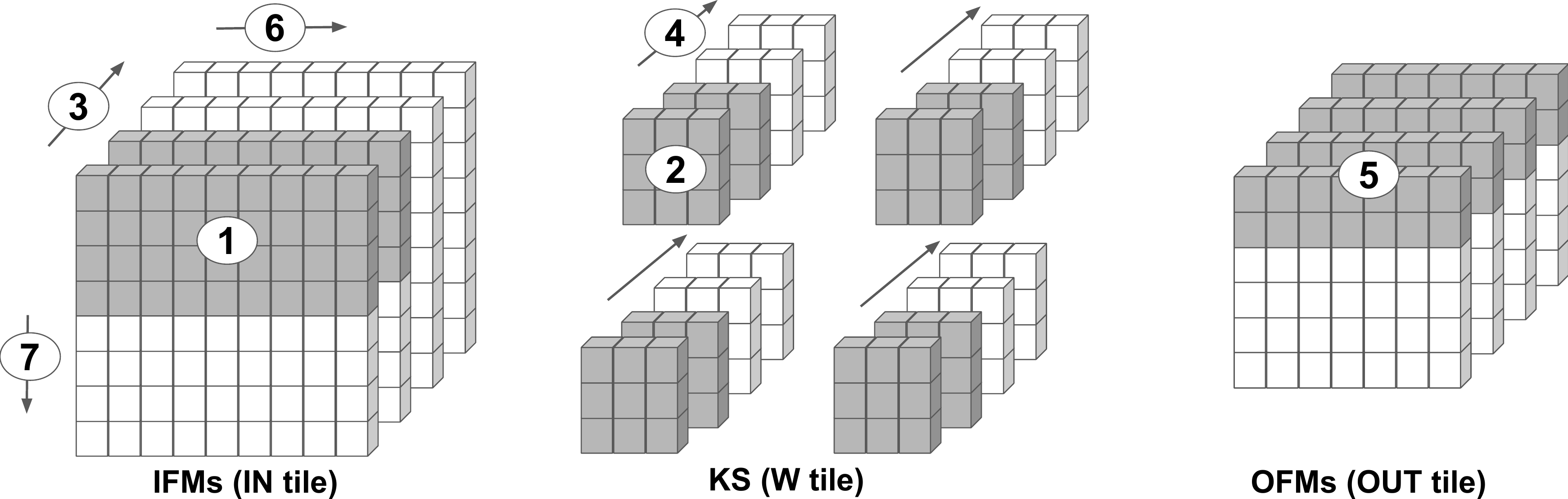}}
  \caption{Input Stationary.}
  \label{fig:input_stationary}
\end{figure}

\textbf{Input Stationary (IS)} -- is a scheduling strategy that focuses on reusing the IN tiles. Figure \ref{fig:input_stationary} shows the execution flow of IS. The first step (\circled{1} in \rfig{input_stationary}) is to load the IN tile from the DRAM into the NMP MB0 on-chip memory; then, the W tile is also loaded \circled{2} from the DRAM into MB1. To make full reuse of the IN tile, the W tile has to include all the filters designated to the TLT -- even if just a small part of each one of them. With the IN and W tiles already loaded, the MAC Unit executes the convolution on them. The result is stored into the MB2 (OUT) memory which, at this point, only contains a partial sum of the Convolution -- the final result of the OUT tile is only generated after computing all elements through the depth of the IFMs. To do that, multiple IN  \circled{3} and W tiles \circled{4} may be required to be loaded while going through the  channel (depth) direction. After computing and accumulating the results, the OUT tile is ready to be stored into the DRAM \circled{5}. After that, a new IN tile is loaded,  going first on the width \circled {6} and then on the height \circled{7} directions of the IFM -- for each one of them, the same W tiles are reloaded again and again.

Given the access pattern performed by IS when loading/storing data from/to the DRAM, one can use Equation \ref{eq:is_alphas} to identify, for each tile (IN, W and OUT tile), the number of times it is required to load/store each one of them to cover the entire computation of a Conv-layer over the TLEs/TLTs. The $\alpha_{in}$ and $\alpha_{w}$ symbols denote the number of times the TLEs/TLTs have to load the IN and W tiles from the DRAM to compute an entire Conv-layer. The $\alpha_{out}$ stands for the number of times the OUT tiles are stored to the DRAM.
\begin{equation}
\centering
\begin{cases}
  \alpha_{in} = \#TLEs \times \ceil*{\frac{TLE_R}{T_R}} \times \ceil*{\frac{C}{T_C}} \times \ceil*{\frac{N}{T_N}} \\
  \alpha_{w} = \#TLEs \times \ceil*{\frac{TLE_R}{T_R}} \times \ceil*{\frac{C}{T_C}} \times \ceil*{\frac{N}{T_N}} \\
  \alpha_{out} = \ceil*{\frac{R}{T_R}} \times \ceil*{\frac{C}{T_C}} 
  \label{eq:is_alphas}
\end{cases}
\end{equation}

\begin{figure}[ht]
  \centering
  \fbox{\includegraphics[scale=0.25]{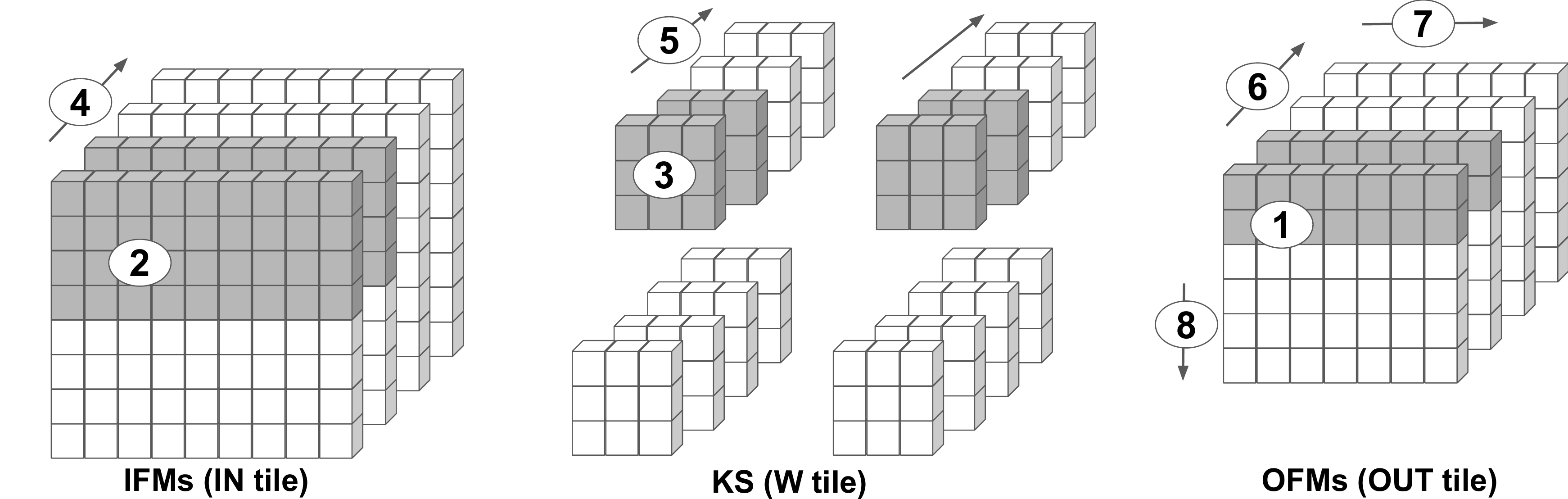}}
  \caption{Output Stationary.}
  \label{fig:output_stationary}
\end{figure}

\textbf{Output Stationary (OS)} -- is a mapping strategy that prioritizes the generation of the OUT tiles, no matter if the same IN and W tiles have to be loaded multiple times from the DRAM into their respective on-chip memories. Figure \ref{fig:output_stationary} shows the execution flow of OS. First, based on the OUT tile \circled{1} dimensions, the corresponding IN \circled{2} and W \circled{3} tiles are loaded from the DRAM into their respective on-chip memories to compute a convolution on them using the TLT's MAC Unit. Given that typically the on-chip memories have not enough space to accommodate all the required input data, multiple IN \circled{4} and  W \circled{5} tiles have to be loaded using the channel (depth) direction.  After finishing the computation of an OUT tile, it is stored into the DRAM and a new OUT tile starts to be computed using the channels' (depth) direction \circled{6}. After that, the other OUT tiles are computed by following first the OFM's width \circled{7}  and then its height \circled{8}. The number of times each IN, W and OUT tile have to be loaded/stored from/to the DRAM is defined in Equation \ref{eq:os_alphas}.
\begin{equation}
\centering
\begin{cases}
  \alpha_{in} = \#TLEs \times \ceil*{\frac{TLE_R}{T_R}} \times \ceil*{\frac{C}{T_C}} \times \ceil*{\frac{N}{T_N} }  \times \ceil*{\frac{TLE_W}{T_M}}  \\
  \alpha_{w} = \#TLEs \times \ceil*{\frac{TLE_R}{T_R}} \times \ceil*{\frac{C}{T_C}} \times \ceil*{\frac{N}{T_N}}  \times \ceil*{\frac{TLE_W}{T_M}} \\
  \alpha_{out} = \ceil*{\frac{R}{T_R}} \times \ceil*{\frac{C}{T_C}}  \times  \ceil*{\frac{M}{T_M}}
  \label{eq:os_alphas}
\end{cases}
\end{equation}

\begin{figure}[ht]
  \centering
  \fbox{\includegraphics[scale=0.25]{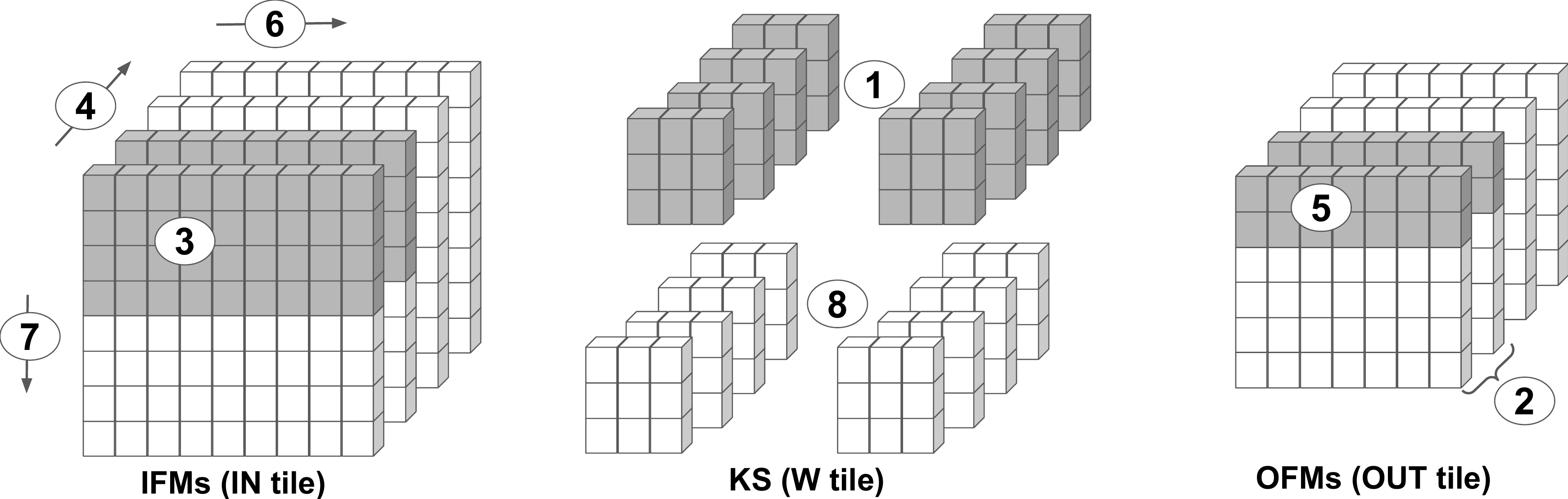}}
  \caption{Weight Stationary.}
  \label{fig:weight_stationary}
\end{figure}

\textbf{Weight Stationary (WS)} -- is a mapping strategy that focuses on loading the filters from the DRAM only once and reuse them as the convolution tiles are computed. Figure \ref{fig:weight_stationary} shows the execution flow of WS. First, the $T_M$ filters in W tile are loaded from the DRAM into the MB1 on-chip memory \circled{1}. In this strategy, each loaded filter includes all its $N$ channels. The loaded filters are then reused until the resulting OFMs are computed \circled{2}. Prior to executing the nmp\_conv2d instruction, an IN tile is loaded from the DRAM \circled{3}. Multiples loads of an IN tile along the channels' depth may be required \circled{4}, each computing and storing partial results that will later be accumulated to form the final OUT tile \circled{5}, so it can be stored back to the DRAM. This is followed by loading other IN tiles in sequence over the width \circled{6} and then over the height \circled{7}. This proceeds until all the OFMs of the respective filters in W tile are computed. After that, a new W tile with other filters may be required to be loaded \circled{8} to compute their corresponding OFMs -- at this point, for each iteration, the same IN tiles are again loaded. Equation \ref{eq:ws_alphas} defines the number of required data transfers to/from the DRAM to cover the entire Conv-layer.
\begin{equation}
\centering
\begin{cases}
  \alpha_{in} = \#TLEs \times \ceil*{\frac{TLE_R}{T_R}} \times \ceil*{\frac{C}{T_C}} \times \ceil*{\frac{N}{T_N}}  \times \ceil*{\frac{TLE_W}{T_M}}  \\
  \alpha_{w} = \#TLEs \times \ceil*{\frac{TLE_W}{T_M}} \\
  \alpha_{out} = \ceil*{\frac{R}{T_R}} \times \ceil*{\frac{C}{T_C}}  \times \ceil*{\frac{M}{T_M}}
\label{eq:ws_alphas}
\end{cases}
\end{equation}

The NMP architecture enables other partitioning strategies beyond the one proposed in this paper, which leverages a MULTICAST instruction to load IFM slices into 8 TLTs/TLE in parallel. For example, one could consider an approach that divides the IFMs (channels) among the TLEs so that they compute partial OFM sums, which are later reduced to the final OFM result. Unfortunately, in NMP, the process of reducing partial sums would require many ring-network messages between the TLEs, thus impacting performance and making some TLTs idle while others are computing the reduce-tree. Moreover, besides leveraging MULTICAST loads to reduce data transfers, the partitioning strategies proposed herein also guarantee load-balancing between the TLEs/TLT.

\subsection{Estimating Time}
\label{sec:burst_modeling}

In order to decide which slicing strategy is the best among those discussed in the sections above, for each convolution TSO combines multiple solutions from the search space $<$$PART_{TLE}$,$PART_{TLT}$,$T_M$,$T_N$,$T_R$,$T_C$$>$, and estimates the time taken by each valid combination to select the one which provides the best performance. This estimate has the following components, listed in increasing order of their contribution to the convolution execution: (a) the time required to run the RISC-V instructions at each TLT; (b) the time needed to perform the MAC unit operations on the slices; and (c) the time required to load/store data between the DRAM and the NPU on-chip memories.  An estimate for the execution time is calculated by function \textit{CalcTime} (defined in \ralg{tlttile} -- line 11), which sums the time of each component of the execution according to Equation \ref{eq:total_time_execution}.
\begin{equation}
\centering
\begin{split}
\large
    T_{CONV} = T_{MAC} + T_{DRAM} + T_{SW}
\label{eq:total_time_execution}
\end{split}
\end{equation}
where $T_{MAC}$, $T_{DRAM}$ and $T_{SW}$ stands for the time taken by the MAC Unit, the time taken to transfer data between the NPU's on-chip memories and DRAM and the time taken to execute the RISC-V instructions, respectively. Since $T_{SW}$ is not significant (usually less than 5\% of the total execution), we will not cover it in detail in this paper.

The time $T_{MAC}$ is calculated according to the number of Multiply-accumulate operations (MAC operations) of a Conv-layer. In the first step (see Equation \ref{eq:numberOfMACs}), it is identified the number of MAC operations required to compute a channel over the IFMs, which is then divided by the number of MAC operations that a MAC unit can execute at each cycle. After that, the other feature maps are then considered to compose the estimated time of the entire tile ($Tile_{MAC}$). Given the time taken of a single tile, it is then possible to estimate the total time required to compute the entire Conv-layer's workload, which is distributed over all the TLTs cores ($\#TLEs \times \#TLTs$) in NMP (see Equation \ref{eq:totalCyclesMac}).
\begin{equation}
\centering
\begin{split}
    Tile_{MAC} = T_N \times T_M \times \left( \ceil*{\frac{T_R \times T_C \times K \times K}{\#MACs}} \right) \times \frac{1}{Freq}
    \label{eq:numberOfMACs}
\end{split}
\end{equation} 
\begin{equation}
\centering
\begin{split}
    T_{MAC} = \frac{1}{\#TLTs} \times \ceil*{\frac{M}{T_M}} \times \ceil*{\frac{N}{T_N}} \times \\ \ceil*{\frac{R}{T_R}} \times \ceil*{\frac{C}{T_C}} \times Tile_{MAC}
\label{eq:totalCyclesMac}
\end{split}
\end{equation} 

To evaluate the data transfer between the on-chip memories and DRAM, TSO uses two approaches to estimate the time taken to load/store the IN, W and OUT tiles. They are: (a) TSO-burst, a burst-based model that estimates the tile's  DRAM transfer time by determining the number of memory bursts and the total CAS and access time it takes; and (b) TSO-noburst, a data volume-based model, which estimates the DRAM transfer time solely based on the size of the tile's data and the memory bandwidth. Algorithm \ref{alg:est_dt} describes how the DRAM transfer time is computed using both approaches. For the sake of explanation of \ralg{est_dt} assume a DDR3 memory with $CAS$ latency (us), $BW$ bandwidth (Bytes/sec) and $BURST$ size in bytes (e.g., 128B)

In order to estimate the time taken by data transfers (see Equation \ref{eq:totalCyclesDRAM}) in both approaches (TSO-burst \& TSO-noburst), the number of memory accesses each tile requires ($\alpha_{in}$, $\alpha_{w}$ and $\alpha_{out}$) is combined with a call to function \textit{CalcDataTransfer} (defined in \ralg{est_dt} -- more details below). The estimated time of each tile is then composed to form $T_{DRAM}$.
\begin{equation}
\centering
\begin{split}
    T_{DRAM} = \alpha_{in} * CalcDataTransfer(IN, conv, arch) + \\ \alpha_{w} * CalcDataTransfer(W, conv, arch) + \\ \alpha_{out} * CalcDataTransfer(OUT, conv, arch)
\label{eq:totalCyclesDRAM}
\end{split}
\end{equation} 

\begin{algorithm}[!t]
\small
\begin{algorithmic}[1]
\caption{Estimate the time taken by Data Transfer}
\Function{CalcDataTransfer}{$tile,conv,arch$}
\State $type \gets arch.type$
\State $BW \gets arch.BW$
\If{$model =$ TSO-burst}
    \State $nbursts \gets$  \Call{CalBurstCount}{$tile, conv, type$}
    \State $cas\_latency \gets nbursts * CAS$
    \State $tile\_size \gets$ \Call{GetTileSize}{$tile, type$}
    \State $transfer\_time \gets tile\_size / BW$
    \State $total\_time \gets transfer\_time  + CAS\_latency$
\Else 
    \LineComment TSO-noburst
    \State $tile\_size \gets$ \Call{GetTileSize}{$tile, type$}
    \State $total\_time \gets tile\_size / BW$
\EndIf
\State \Return $total\_time$
\EndFunction
\label{alg:est_dt}
\end{algorithmic}
\end{algorithm}

\textbf{Burst-based data transfer (TSO-burst)} -- The key idea behind TSO-burst is to determine the number of bursts taken by each access to a tile row to use it to determine an estimate for the DRAM access time of the tile. For instance, consider  an IN tile containing $T_N$ (channels) $\times$ $T_H$ (rows) $\times$ $T_L$ (columns) where each entry has 16-bit (2B) elements. Given that the channel is laid out in row-major, loading the first element of a row takes time  $CAS$, while loading the remaining elements $\sim (2\times (T_L - 1))/BW$. Thus, an IN row takes $CAS + \sim (T_L-1)/BW$ to load. This is true if the size of the row ($2\times T_L$) is smaller than $BURST$ bytes. Otherwise, other memory bursts may occur when loading the row, and additional $CAS$ penalties will impact the time. 

Algorithm \ref{alg:est_dt} is used to estimate the execution time when convolution $conv$ is divided into tiles $tile$ on architecture $arch$. Initially, the tile data size $type$ (e.g., 16-bit fixed-point)  (line 2) and the DRAM memory bandwidth $BW$ (line 3) are extracted from the $arch$ data structure. Next (line 4), the algorithm selects the memory transfer model (e.g., TSO-burst) and uses a call to function \textit{CalBurstCount} (line 5) to determine the number of bursts ($nburts$) required to load all the $T_H$ rows of an (IN) tile. Then, the impact of the CAS latency is computed into $cas\_latency$ (line 6) and the size of the tile ($tile\_size$) is determined in line 7 by calling function \textit{GetTileSize}. The time to transfer all the data in a tile ($transfer\_time$) is then determined (line 8), and finally, the total time to load the tile is computed (line 9) and returned (line 14).

TSO-burst does not make any assumptions about the external DRAM or memory-controller designs, besides the existence of burst-based accesses typically found in these memories. The memory-controller found in the NMP board follows the ARM-bus protocol. Besides CAS-latency, other DRAM parameters (e.g., Trcd/Trp/Tras)  could also be included to improve the precision of data transfer modeling. Nevertheless, since  CAS-latency is the most relevant of these DRAM parameters \ralg{est_dt} focused only on it.

\textbf{Volume-based data transfer (TSO-noburst)} -- This approach  is typically used by all previous works which address this problem. As  shown in lines 12-13 of Algorithm \ref{alg:est_dt}, it estimates the tile time by considering only the time to transfer the tile data (line 12) and not the impact of the CAS latency of the tile's memory bursts.

\section{NMP XLA Compiler} 
\label{sec:tf_xla}

In this work, we  used Tensorflow XLA (TF-XLA) \cite{tfxla} which is a domain-specific compiler for linear algebra. TF-XLA Ahead-of-time (AOT) compilation was used to generate executable binaries for machine learning models on the NMP architecture.

\begin{figure}[!t]
  \centering
  \includegraphics[scale=0.6]{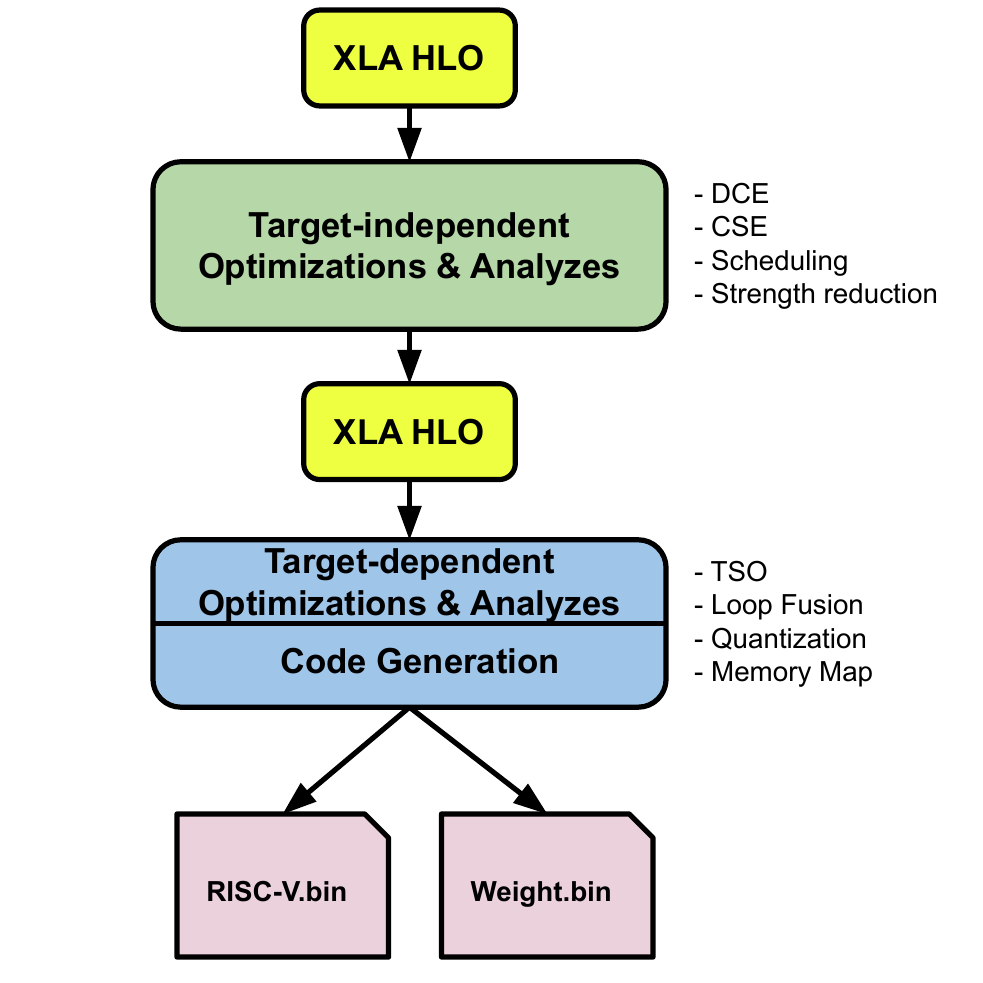}
  \caption{Tensorflow XLA Flow.}
  \label{fig:xla_flow}
\end{figure}

\textbf{Compilation flow} -- TF-XLA compiler receives as input a \textit{protobuf} file, that contains the definition of the network (operations and their connections) and the weights. It then performs some transformations (using the TF Graph Transform Tool) on the \textit{protobuf}, (e.g., folding batch-norm into convolutions), which are useful later during the quantization pass. After the transformations are applied, a new protobuf is generated, which is then used to create the XLA HLO intermediate representation (see Figure \ref{fig:xla_flow}). From the initial XLA HLO representation,  target-independent optimizations are applied (e.g., DCE, CSE, etc), thus producing an optimized HLO representation. After that, target-dependent optimizations, like  quantization, are executed followed by other specific passes to ease code generation (e.g., operation fusion). TSO  is the last pass applied to the HLO IR, just before LLVM IR lowering. After that, the compiler generates LLVM IR from HLO IR, which calls NMP intrinsics to lower computation to TLT RISC-V accelerated code.

\textbf{Quantization pass} -- in this pass, our TF/XLA compiler converts weights from 32-bit floating point to 8- or 16-bit fixed-point. The goal of quantization is to determine the \textit{Q-points} for the input, intermediate tensors and weights. The Q-point is a radix point that determines how many bits are assigned to the integer and fractional parts of the data. The goal is to ensure that the integer part has enough bits to represent the minimum and maximum output values that an operation  produces. To capture the data range of the model tensors some steps need to be performed. First, one has to identify the minimum and maximum values each layer of the CNN can produce for a set of different input tensors. To achieve that,  the \textit{insert\_logging} $method$ available in the TF Graph Transform Tool is used to insert probes at the inputs/outputs of all model operations, producing an instrumented \textit{protobuf}.  After that, the compiler executes a calibration step, which consists on the execution of the instrumented \textit{protobuf} over hundreds or thousands of images to capture the minimum and maximum values that each logged operation outputs.  These values are then used to determine the Q-point for each of the logged operations. Given that some operations are not marked during the process of logging (e.g., Pool-Layer), a sequence of two traversals is performed on the TF Graph. In the first traversal (forward), the already computed Q-points are propagated to the operators from the input towards the output of the graph.  In the second traversal (backward), from the output of the graph to its input, the Q-points are adjusted -- for example, the inputs of a Concatenate or an Add operation have to have the same Q-points. The compiler then modifies the Layout from NHWC (default on Tensorflow) to NCHW, which is the layout required by the NMP architecture, and then generates a file with the quantized weights. Besides that, a memory map is also generated, which is used at run time by the binary code of each TLT to schedule tiles so they can load/store data in the DRAM and TLT on-chip memories.

After optimizing the HLO IR, our TF-XLA compiler lowers the HLO IR to LLVM IR. During code generation, the  HLO instructions map to intrinsics in an optimized NMP library, which includes most of the typical CNN operations. From the LLVM IR, the compiler generates a RISC-V executable (RISCV.bin) that is used by the TLTs, together with the quantized weight file (Weight.bin), to execute the model.

Other AI compilers have also been used to generate code for AI accelerators. As an example, Glow \cite{jin2020compiling} generates code for Intel Habana, and onnx-mlir \cite{rotem2018glow} generates code for an AI accelerator integrated into the IBM z16 processor.

The execution of TSO takes a couple of minutes. In order to easy retargetability, we intend in the future to make TSO a generic MLIR-pass to be used by flows like ONNX and XLA in a machine-independent manner. The user will specify the memory-hierarchy, the number of processors/cores and the MLIR-TSO pass will output the best tiling/scheduling scheme so LLVM can lower it.

\section{Experimental Results}
\label{sec:results}

In order to validate the TSO approach, a set of experiments was executed on an NMP board equipped with 4 TLEs, each having 8 TLTs. Each TLT contains three 8KB MB on-chip memories (MB0--MB2). To evaluate TSO, we used used 5 CNN image classification models:  InceptionV3 \cite{szegedy2016rethinking}, LeNet \cite{lecun1998gradient}, MobileNetV2 \cite{sandler2018mobilenetv2}, ResNetV50 \cite{he2016deep}, and SqueezeNet \cite{iandola2016squeezenet}. We have also applied TSO to an object detection application - a YOLO-based model \cite{silva2017real} used to recognize car license plates. The selected models have a varied number of convolutions with different shapes of input (IFMs), weight (KS) and output (OFMs).

\begin{table}[!t]
\resizebox{\columnwidth}{!}{%
\begin{tabular}{|l|cccc|cc|ccc|ccc|}
\hline
\multirow{3}{*}{\textbf{Model}} & \multicolumn{4}{c|}{\textbf{Accuracy (TF-XLA)}} & \multicolumn{2}{c|}{\multirow{2}{*}{\textbf{TSO (us)}}} & \multicolumn{3}{c|}{\multirow{2}{*}{\textbf{\begin{tabular}[c]{@{}c@{}}TLE\\ partitioning\\ fixed (us)\end{tabular}}}} & \multicolumn{3}{c|}{\multirow{2}{*}{\textbf{\begin{tabular}[c]{@{}c@{}}TLT\\ scheduling\\ fixed (us)\end{tabular}}}} \\ \cline{2-5}
 & \multicolumn{2}{c|}{\textbf{\begin{tabular}[c]{@{}c@{}}CPU FP \\ (\%)\end{tabular}}} & \multicolumn{2}{c|}{\textbf{\begin{tabular}[c]{@{}c@{}}NMP \\ (\%)\end{tabular}}} & \multicolumn{2}{c|}{} & \multicolumn{3}{c|}{} & \multicolumn{3}{c|}{} \\ \cline{2-13} 
 & \multicolumn{1}{c|}{\textbf{Top-1}} & \multicolumn{1}{c|}{\textbf{Top-5}} & \multicolumn{1}{c|}{\textbf{Top-1}} & \textbf{Top-5} & \multicolumn{1}{c|}{\textbf{Burst}} & \textbf{\begin{tabular}[c]{@{}c@{}}No\\ Burst\end{tabular}} & \multicolumn{1}{c|}{\textbf{KS}} & \multicolumn{1}{c|}{\textbf{\begin{tabular}[c]{@{}c@{}}KS\&\\ OFM\end{tabular}}} & \textbf{OFM} & \multicolumn{1}{c|}{\textbf{IS}} & \multicolumn{1}{c|}{\textbf{OS}} & \textbf{WS} \\ \hline
InceptionV3 & \multicolumn{1}{c|}{75.1} & \multicolumn{1}{c|}{92} & \multicolumn{1}{c|}{76.9} & 93.4 & \multicolumn{1}{c|}{72686} & 88478 & \multicolumn{1}{c|}{82370} & \multicolumn{1}{c|}{81367} & 91408 & \multicolumn{1}{c|}{73706} & \multicolumn{1}{c|}{93731} & 101641 \\ \hline
LeNet & \multicolumn{1}{c|}{99.9} & \multicolumn{1}{c|}{100} & \multicolumn{1}{c|}{99.9} & 100 & \multicolumn{1}{c|}{199} & 231 & \multicolumn{1}{c|}{202} & \multicolumn{1}{c|}{224} & 231 & \multicolumn{1}{c|}{199} & \multicolumn{1}{c|}{233} & 228 \\ \hline
MobileNetV2 & \multicolumn{1}{c|}{70.2} & \multicolumn{1}{c|}{89.6} & \multicolumn{1}{c|}{70.5} & 89.8 & \multicolumn{1}{c|}{14030} & 15470 & \multicolumn{1}{c|}{17765} & \multicolumn{1}{c|}{16696} & 17571 & \multicolumn{1}{c|}{14084} & \multicolumn{1}{c|}{17387} & 15155 \\ \hline
ResNet-50 & \multicolumn{1}{c|}{70.6} & \multicolumn{1}{c|}{89.9} & \multicolumn{1}{c|}{70.7} & 89.9 & \multicolumn{1}{c|}{55927} & 62375 & \multicolumn{1}{c|}{62077} & \multicolumn{1}{c|}{63118} & 78844 & \multicolumn{1}{c|}{59714} & \multicolumn{1}{c|}{71905} & 77535 \\ \hline
SqueezeNet & \multicolumn{1}{c|}{47.1} & \multicolumn{1}{c|}{71} & \multicolumn{1}{c|}{47.1} & 71 & \multicolumn{1}{c|}{12504} & 13713 & \multicolumn{1}{c|}{16579} & \multicolumn{1}{c|}{14134} & 12993 & \multicolumn{1}{c|}{12908} & \multicolumn{1}{c|}{15840} & 13452 \\ \hline
YOLO & \multicolumn{1}{c|}{-} & \multicolumn{1}{c|}{-} & \multicolumn{1}{c|}{-} & - & \multicolumn{1}{c|}{53271} & 57225 & \multicolumn{1}{c|}{56591} & \multicolumn{1}{c|}{55713} & 63550 & \multicolumn{1}{c|}{68530} & \multicolumn{1}{c|}{58592} & 55375 \\ \hline
\end{tabular}}
\caption{Model accuracy on CPU (FP32) and NMP (16-bit fixed point), execution times for TSO (burst and noburst) and fixed strategies. YOLO uses a different metric for accuracy, it measures the precision of the detection, which is 93.53\% on NMP while on CPU is 93.03\%.}
\label{tab:timesandccuracy}
\end{table}

All models were compiled by our TF-XLA compiler with the quantization pass set to 16-bit fixed-point. The accuracy achieved by each of the image classification models on NMP is shown in Table 1. The Top-1 and Top-5 accuracies were measured by running all images from the validation datasets, MNIST~\cite{mnist} and ImageNet (ILSVRC2012)~\cite{imagenet}, for LeNet and the other image classification models, respectively. The same datasets were also used to measure the original floating-point models (FP 32-bit) on CPU. The difference in terms of accuracy drop ranges from 0.1 up to 1.8\%. For the YOLO-based model, NMP reaches a precision of detection of 93.53\% on a car plate dataset~\cite{gonccalves2016benchmark} executed on 16-bit quantized data, and the same model on CPU results in 93.03\%.

The quantization scheme used in this paper only quantizes the convolution data to 16-bit fixed-point. But, for smaller precision, e.g., 8-bit fixed-point, TSO may decide on selecting a different solution than the one it would select for the same convolution quantized to 16-bit fixed-point. For instance, assume that only after quantizing a given convolution to 8-bit, one of the components of the convolution (e.g., IFMs) becomes smaller than its corresponding MBLOB. Keeping this component stationary would result in a single load of it and of the other components' tiles (KS and OFMs) to compute the convolution. As a result, it would reduce considerably the data transfer between DRAM and NMP. Therefore, TSO would tend to select this solution as it would reduce data transfer over the others.

\subsection*{TSO-burst vs TSO-noburst}

\begin{figure}[!t]
  \centering
  \includegraphics[scale=0.30]{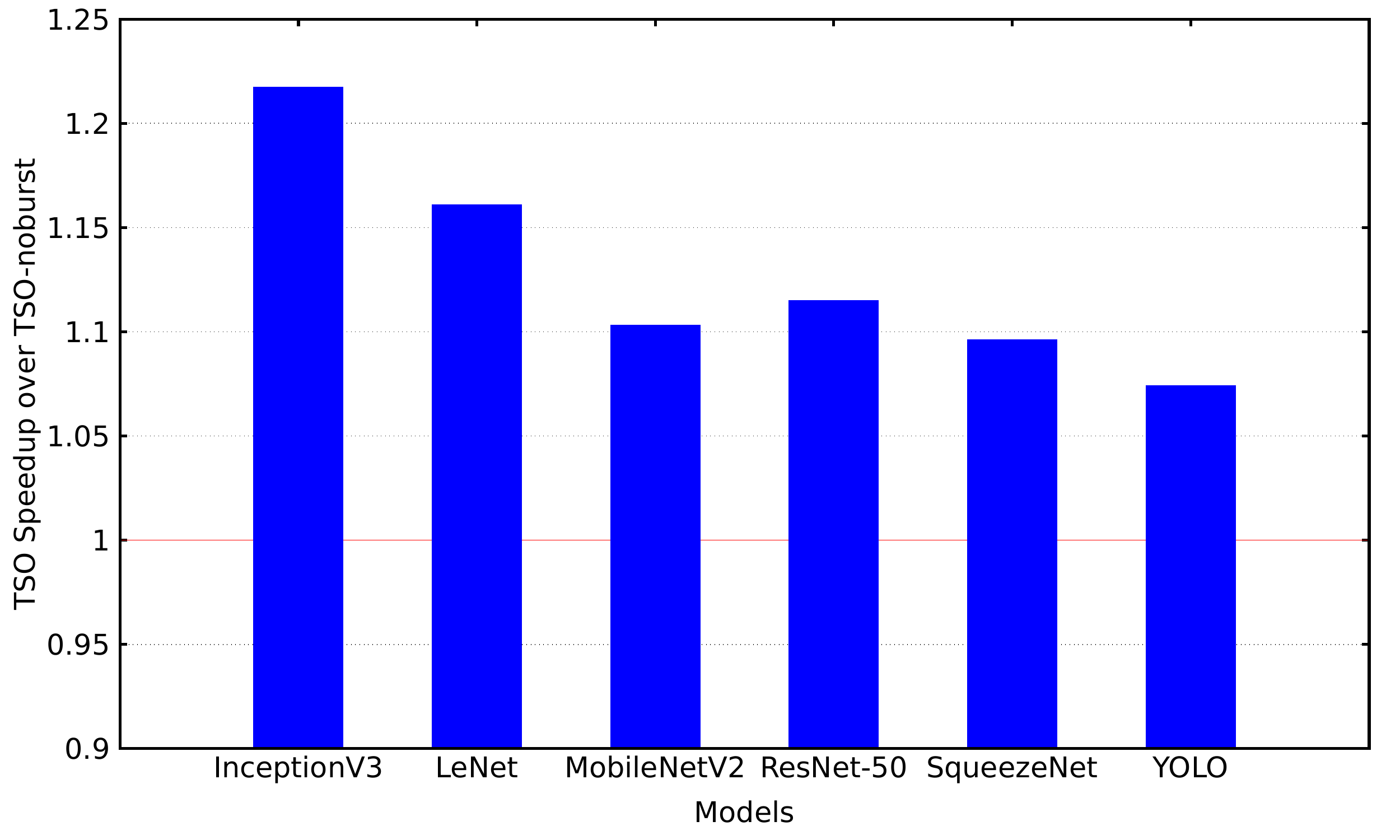}
  \caption{TSO-burst speedup over TSO-noburst.}
  \label{fig:res_burst}
\end{figure}

We compared TSO with the burst-based modeling activated (TSO-burst) and without it (TSO-noburst), as the latter is a common approach found in most previous works. The speedup of TSO-burst over TSO-noburst ranges from 7.4\%, for YOLO, up to 21.7\%, for InceptionV3, as shown in Figure \ref{fig:res_burst}. The main improvements from using TSO-burst come when the IFMs are divided into IN tiles. This happens because  TSO tends to select larger tiles on the width (row-major) direction. By selecting larger tiles, TSO minimizes the number of required bursts, thus reducing the impact of the CAS latency on the memory access time. By prioritizing bursts on the width direction, TSO maximizes the usage of the bursts, as it improves memory access coalescing.  For the case of TSO-noburst, the tiles are selected so as to reduce the number of bytes loaded from the DRAM to NMP. This approach is adopted by most solutions that have been  proposed so far in the literature~\cite{hu2019resources,tu2017deep,zhang2015optimizing,motamedi2016design}. Contrary to those, the TSO-burst technique proposed in this paper takes into consideration DRAM access coalescing to estimate the time taken to LOAD/STORE data from memory, thus resulting in better partitioning and improved performance. The resulting execution time for the various models is shown in Table \ref{tab:timesandccuracy}. Notice in the table that TSO always produces the shortest execution.

As an example, consider the 5th Conv-layer of InceptionV3, which has 80 IFMs of size 73x73 each, and 192 filters of size 80x3x3. The shape of the IN tile selected by TSO-burst has size 14x4x73 ($T_N \times T_H \times T_L$). Since the IN tile of the TSO-burst takes the whole width $L$ (73) of a channel, it results in a sequence of 584 bytes aligned sequentially on the DRAM ($4 \times 73 \times 2B$), which requires 5 memory bursts for each tile's channel. In total, when considering the 14 channels of that tile, the 5th Conv-layer requires a total of 70 memory bursts. On the other hand, TSO-noburst selects a tile of size 16x11x20, which corresponds to only 20 bytes aligned on the DRAM, thus resulting in one memory burst for each row. Given that the IN tile has 16 channels, each with 11 rows, it requires a total of 176 memory bursts, which is more than double what is needed to load the TSO-burst tile.  For that specific 5th  Conv-layer, TSO-burst reduces the tile execution time by 28\%.

\begin{figure}[!t]
  \centering
  \includegraphics[scale=0.25]{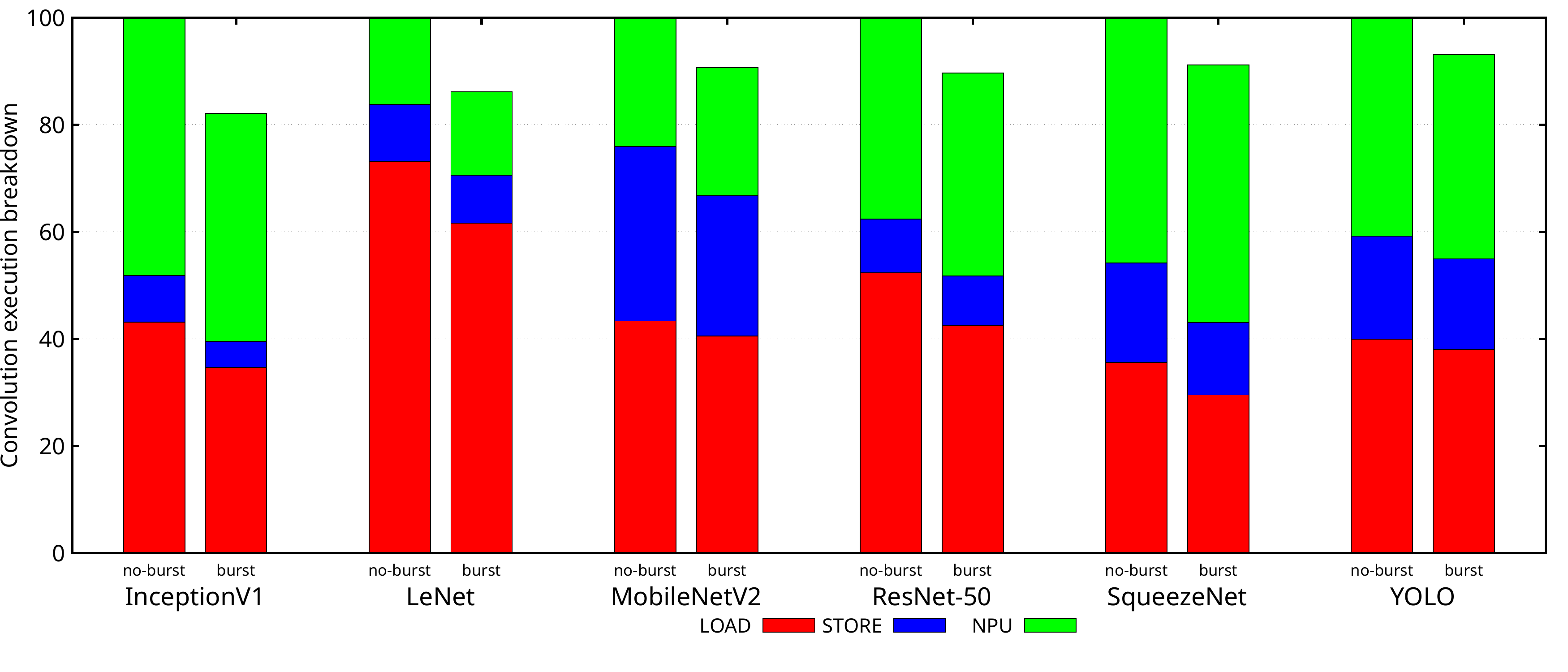}
  \caption{Convolution execution breakdown with TSO-burst as a relative proportion of TSO-noburst.}
  \label{fig:breakdown}
\end{figure}

The solution provided by TSO-burst aims to reduce the total time taken for data transfer operations. To illustrate that, refer to Figure \ref{fig:breakdown}, which shows for each model two bars representing the breakdown of the percentage of computation time spent in LOAD, STORE, and MAC operations with respect to the total execution time. For the TSO-burst's bars, the percentage of the execution time is calculated with respect to the  TSO-noburst total time. As shown in \rfig{breakdown}, when TSO-noburst is used, the percentage of the LOAD+STORE transfer time ranges from 51.83\%, for InceptionV3, up to 83.80\% for LeNet. On the other hand, when  TSO-burst is used to model memory access during TF-XLA compilation, the time taken by LOAD+STORE operations decreases from 7.08\% to 23.71\% for YOLO and InceptionV3, respectively.

\subsection*{Speeding-up TSO solution exploration}

\begin{figure}[!t]
  \centering
  \includegraphics[scale=0.31]{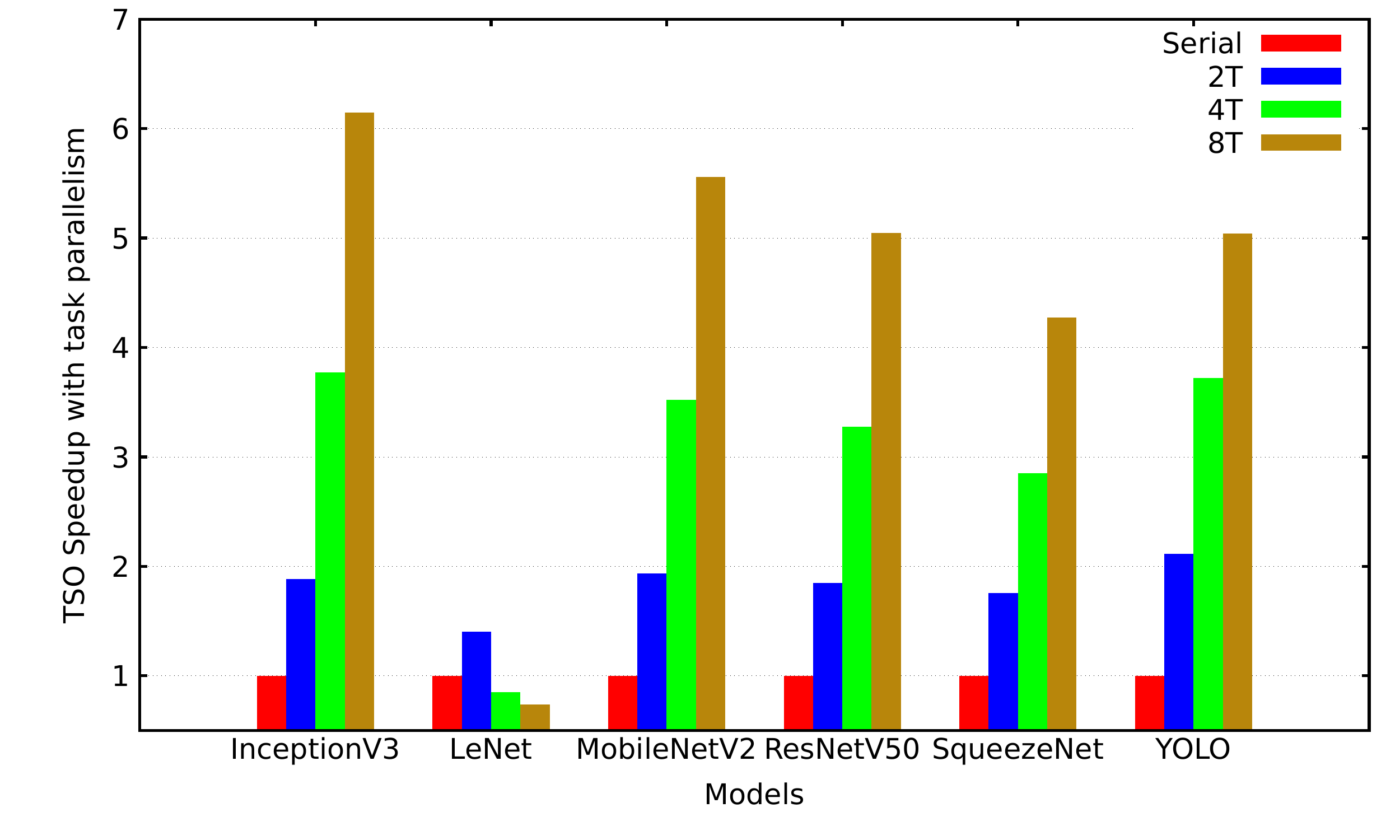}
  \caption{Evaluating OpenMP parallelization of TSO solution space exploration.}
  \label{fig:task_parallelism}
\end{figure}

The goal of this experiment is to evaluate the impact of the OpenMP task-parallelism annotations in \ralg{sel_conv_mapping} (lines 2, 3, and 5) on the overall time of the TSO slicing space exploration.  We did this experiment on an Intel Xeon E5-2620 with 16-physical cores and 64GB of memory. The results are shown in Figure \ref{fig:task_parallelism} where each line corresponds to one model, the y-axys is speedup with respect to sequential execution as the number of threads used by OpenMP grows (x-axys). Notice that the multi-threading execution has almost a linear improvement when compared to the serial execution for most of the models. For InceptionV3, which has 94 Convolutions, the multi-threading execution is almost linear. On the other hand, for LeNet, which has only 2 Convolutions, 2 threads are enough to accelerate the execution, and thus a  slow down shows up if the number of threads increases from that point on. In terms of time, the serial execution of TSO varies from 28 ms, for the LeNet network, to 6 min for the InceptionV3 network, while with the multi-threading execution, this time is reduced to 17 ms and 59 sec, respectively. The total time spent by the compiler, from the beginning to the generation of the binaries, varies from 20 seconds to 7 minutes, for the same models, respectively, whereas most of the time is consumed by the calibration/quantization step.

\begin{figure}[!t]
  \centering
  \includegraphics[scale=0.18]{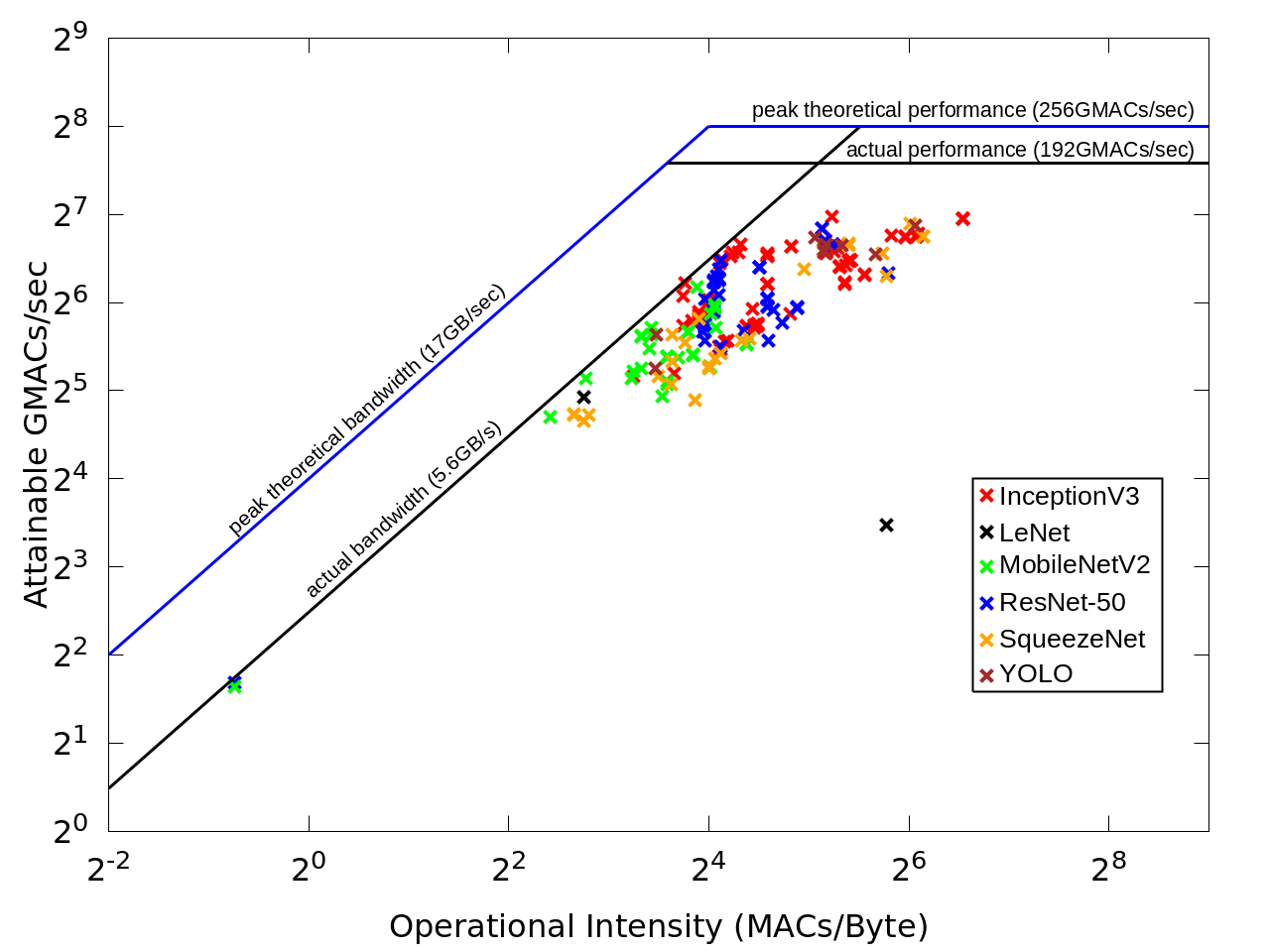}
  \caption{Roofline model for NMP architecture (TSO-burst).}
  \label{fig:roofline_model}
\end{figure}

\subsection*{Roofline Model}

To evaluate the performance of the code resulting from using TSO  on the Conv-layers of  each model executed on NMP (TSO-burst), we used the Roofline Model shown in Figure \ref{fig:roofline_model}. In the graph, the  y-axis presents the Multiply-and-accumulate (MAC) throughput (in GMACs/sec) achieved by the architecture and the convolution execution. In the x-axis is the Operational Intensity, which stands for the number of MAC operations executed for each byte that is loaded from the DRAM. The blue lines in the graph represent the theoretical roofs for both the MAC throughput (horizontal line)  and DRAM bandwidth (sloped line) that can be respectively achieved by the NMP engine and the memory system. To better evaluate the real performance of the system two additional experiments were undertaken to measure these parameters. This is required given that other architecture components can impact their values. The black lines in the graph represent these measurements. As shown, the measured MAC throughput reaches a roof of 192 GMACs/sec, represented by the horizontal black line. A number of issues can explain this reduction. For example, the NMP device used in this work has single-ported on-chip memories, and thus  TLTs that are waiting for data get idle without using its MAC Unit. As for the  memory bandwidth (the sloped black line) the measured value is also reduced. This  can be explained by the fact that the DRAM bandwidth is constrained by a single DMA engine per TLE which has to simultaneously serve all 8 TLT cores.  In order to evaluate the performance resulting from TSO, we plotted one point in Figure \ref{fig:roofline_model} for all convolutions in the models. As shown, most of the convolutions reach either the roof limited by the (measured) memory bandwidth  (sloped black line), or approach the roof defined by the MAC throughput (horizontal black line). This makes it clear that TSO produces code which approaches the maximum performance of the architecture. 

\subsection*{Fixed TLE/TLT Partitioning}

\begin{figure}[t]
  \centering
  \includegraphics[width=1\linewidth]{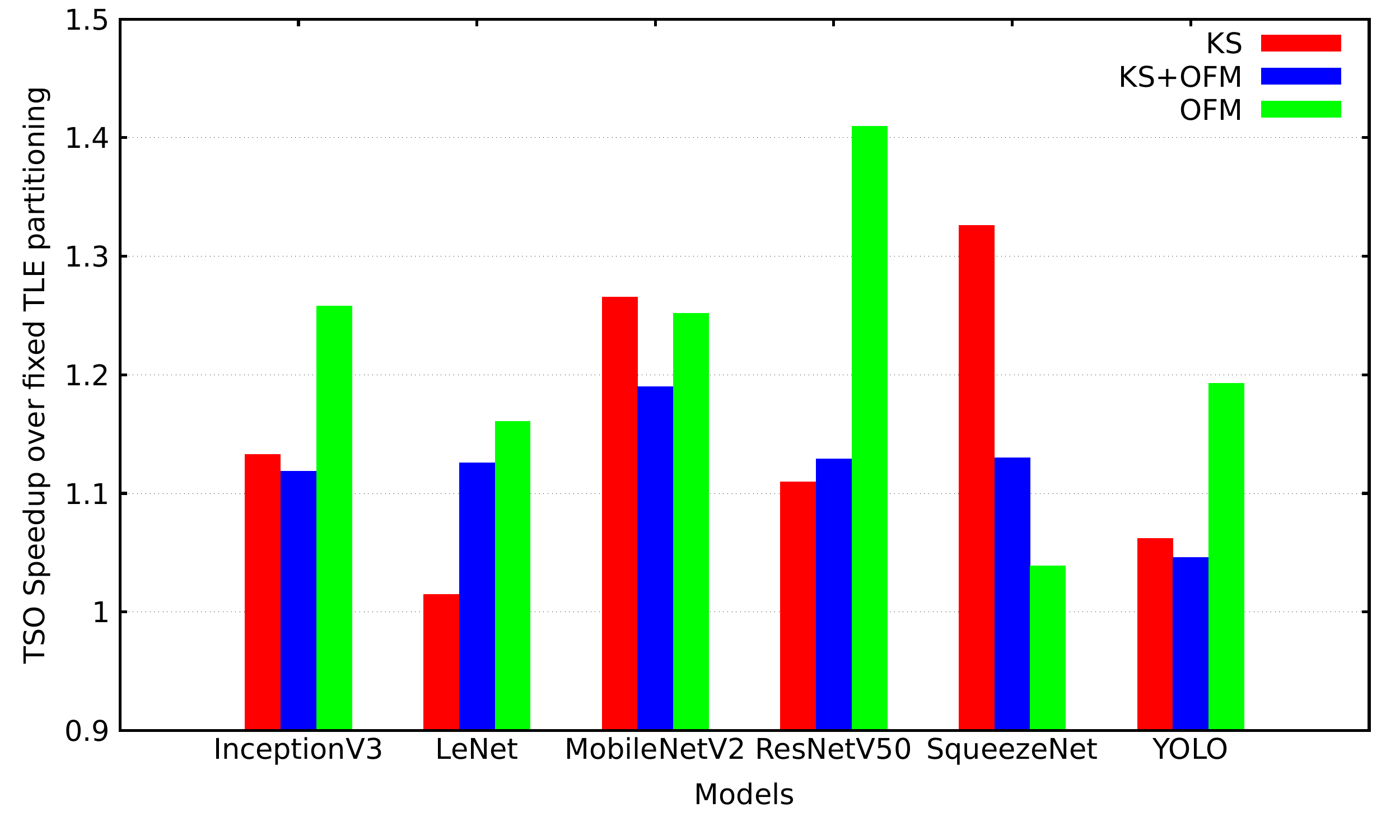}
  \captionof{figure}{Fixed TLE Slicing.}
  \label{fig:tle_part}
\end{figure}

\textbf{Fixed TLE partitioning} -- in this experiment, the compiler was set to generate code which fixes each TLE slicing strategy described in Subsection \ref{sec:TLEPart} for all Conv-layers of a model. The experiment works as follows. The compiler identifies, for the fixed TLE slicing, the best TLT tiling/scheduling strategy (IS, OS and WS). The result of this experiment is shown in Figure \ref{fig:tle_part} which reports the speedup of the model compiled with TSO (burst mode) when compared to the model compiled with the fixed TLE slicing. For the \textit{KS} case,  TSO achieves a speedup of up to 32.6\%, for SqueezeNet.  SqueezeNet does not perform well for TLE slicing since most of its Conv-layers have  IFMs larger than the size of the filter set (KS). For \textit{KS$\&$OFM}, TSO speedup reaches up to 19\%, for MobileNetV2. This TLE strategy usually works better for the Conv-layers that have similar sizes for both IFMs and weights (KS). For \textit{OFM}, TSO speedup is 41.0\%, for ResNet-50. For most of ResNet-50's  Conv-layers, the size of the weight set KS is larger than the size of the IFM data maps. As a result, TSO outperforms the best fixed TLE slicing strategy.

\begin{figure}[t]
  \centering
  \includegraphics[width=1\linewidth]{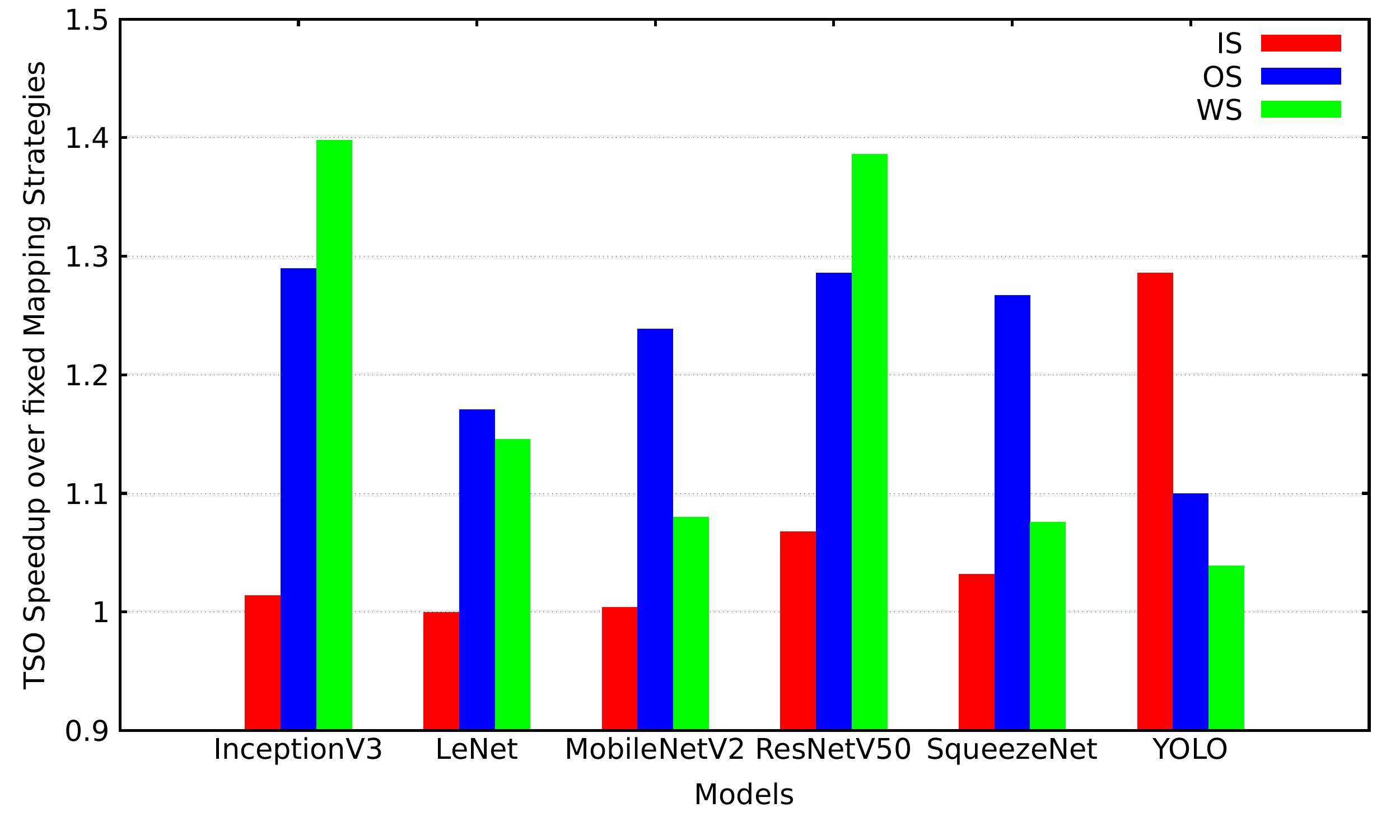}
  \captionof{figure}{Fixed TLT Scheduling Strategy.}
  \label{fig:mapping_strategy_res}
\end{figure}

\textbf{Fixed TLT partitioning} -- in this experiment, the compiler is set to generate code which fixes one of the three TLT scheduling strategies (IS, OS, and WS) for all the model's Conv-layers. The compiler applies the fixed scheduling strategy to all possible TLE slicing options (KS, KS$\&$OFM and OFM) to search for the best performance. Figure \ref{fig:mapping_strategy_res} shows the speedup of TSO when compared to the best fixed TLT strategy. By fixing IS during TF-XLA Code Generation, TSO speedup reaches 28,6\% on YOLO. For the YOLO network, IS does not perform well since this network has multiples Conv-layers with IFMs varying from $102 \times 102$ ($H \times L$) to $416 \times 416$ -- this results in multiples loads of the filters since multiple IN tiles are required. TSO speedup with respect to fixed OS reaches up to 28.6\% on ResNet-50. When compared to fixed WS, TSO speedup is 39.8\% on InceptionV3. In the case of InceptionV3, when WS is used, multiples W tiles are required to work on the slices, and thus multiples loads of the IFMs become necessary for each W tile, leading to an increase in data transfers. Here again,  TSO outperforms the best fixed TLT scheduling strategy.

\subsection*{Evaluation on the Glow Framework}
A set of experiments were also performed to evaluate the portability of TSO to another Machine Learning Framework. The Glow toolchain from Facebook was selected for this evaluation, and a performance comparison was done with respect to the accuracy resulting from the TensorFlow/XLA compiler. A thorough performance comparison with respect to TensorFlow/XLA was not done given that Glow uses a slightly different set of models from the ONNX Model Zoo. 

\begin{table}[]
\centering
\footnotesize
\begin{tabular}{|l|c|c|c|}
\hline
\multicolumn{1}{|c|}{\textbf{Model}}  & \textbf{Arch.} & \textbf{Top-1} & \textbf{Top-5} \\ \hline
\multirow{2}{*}{MNIST 1.3} &  CPU (FP) & 98.9\% & 100\% \\ \cline{2-4} & NMP & 99.2\% & 99.8\% \\ \hline
\multirow{2}{*}{LeNet} & CPU (FP) & 94.8\% & 99.9\% \\ \cline{2-4}  & NMP & 95.2\% & 99.9\% \\ \hline
\multirow{2}{*}{Resnet18 1.2.1} & CPU (FP) & 69.9\% & 89.3\% \\ \cline{2-4} & NMP & 66.4\% & 87.3\% \\ \hline 
\multirow{2}{*}{SqueezeNet 48.7} & CPU (FP) & 49.0\% & 72.9\% \\ \cline{2-4} & NMP & 47.1\% & 71.0\% \\ \hline
\multirow{2}{*}{Mobilenet 2.1} & CPU (FP) & 71.8\% & 90.6\% \\ \cline{2-4} & NMP & 70.9\% & 89.4\% \\ \hline
\end{tabular}
\caption {Model accuracy on CPU (FP32) and NMP (16-bit fixed point) from Glow.}
\label{tab:glowaccuracy}
\end{table}

Some models were required to be converted from the Google TFLite Hub to the ONNX format (e.g., Lenet, Squeezenet, and Mobilenet) while others were used directly from the ONNX Model Zoo (e.g., MNIST and Resnet18). Accuracies have been compared to those achieved on CPU 32-bit FP, as listed in the corresponding repositories. To achieve that, inferences for each model were executed on NMP architecture over the ImageNet and MNIST validation datasets, according to the model, and the Top-1 and Top-5 accuracies were measured. As shown in \rtab{glowaccuracy}, accuracies of the code produced by Glow on NMP approach those from the TensorFlow/XLA compiler (see Table \ref{tab:timesandccuracy}).

\section{Related Works}
\label{sec:related_works}

Maestro \cite{kwon2020maestro} and Timeloop \cite{parashar2019timeloop} uses analytical modeling that evaluates different mapping configurations -- dataflow strategy, data-reuse, tile size, etc; to estimate the runtime for different configurations. While Maestro designs some annotations to classify the loops either as temporal or spatial, Timeloop analysis the nested loops to apply the transformations on them. For both, given a DNN layer (e.g., a Convolution and its information), hardware configuration (number of PEs, on-chip memories size, etc), the dataflow strategy, these approaches estimate the runtime performance, energy and power. Given the easy use of Maestro, different solutions have adopted its annotations to estimate computation \cite{chatarasi2020marvel,kao2020gamma}. As an example, Marvel \cite{chatarasi2020marvel} uses the Maestro notations and has for  main goal the reduction of the search space by decoupling the analysis of the cost model of the accesses to the on-chip/off-chip sub-spaces. Timeloop and Maestro model Spatial DNN Accelerators, i.e., FPGA-based architectures, in which the inner-loops of a Convolution are unrolled and then synthesized into PE array (MAC units) which run in a synchronized fashion to leverage on data-sharing between them through inter-PE communication. Similar to Marvel and Timeloop, our work also performs cost modeling and design space exploration, but  contrary to them, we model execution on multicore NPU architectures and not on FPGA designs.

Tu \textit{et al.} \cite{tu2017deep} and Hu \textit{et al.} \cite{hu2019resources} proposed an FPGA-based accelerator capable of reconfiguring its resources to increase data reuse. They used the concept of Input Stationary (IS), Weight Stationary (WS), and Output Stationary (OS). Besides that, they propose a novel approach called Hybrid Stationary (HS) that leverages on these concepts to find an optimal configuration for each Conv-layer. Although their work has some similarity to ours, instead of mapping the operations to an array of PEs, we consider an architecture (NMP) with multiple cores where each core has an accelerator which runs independently of the others. Besides that, our search space exploration algorithm considers different tile shapes based on memory bursts, and not just square shapes that fit into the hardware topology.

To select different tile sizes, loop order, unroll factor, etc, TVM \cite{chen2018tvm} uses a machine-learning cost model, which does not require hardware information, and periodically learns from previous predictions to search for an improved partitioning. To the best of our knowledge, and from the available public literature, TVM has not shown any results for multicore NPUs like NMP, generating code only for FPGAs, embedded CPUs and server CPUs.

To improve data reuse, some works use polyhedral-based optimization techniques \cite{peemen2013memory,zhang2015optimizing}. Ma \textit{et al.} \cite{ma2019performance} describes a performance model that implements Output Stationary (OS). Chen \textit{et al.} \cite{chen2016eyeriss} describes an approach called Row Stationary (RS) that minimizes data movement by exploiting data reuse through inter-PE communication. The tile selection on those works is usually selected from the use of a roofline-based model \cite{motamedi2016design,park2020roofline}. Compared to our work, their roofline model only considers the number of memory accesses without taking memory burst into consideration. Stoutchinin \textit{et al.} \cite{stoutchinin2019optimally}, on the other hand, uses a technique called reuse distance, which aims to identify the memory footprint which is required to accommodate the Convolution's data into the on-chip memory, which varies up to 512KB. His work only considers data reuse over the on-chip memories without taking into consideration DRAM accesses.

Caffeine \cite{zhang2018caffeine} is a library that comes with the capability of converting Fully Connected Layers (FCL) into Conv-Layer. The conversion takes into consideration modifications in the data-layout to reduce the number of accesses to the DRAM so as to increase the burst length. Qiu \textit{et al.} \cite{qiu2016going} also modifies the data layout and applies quantization to improve memory access. Putra \textit{et al.} \cite{putra2020drmap} maps the data in the DRAM to reduce row buffer conflicts. Our work uses a similar idea to increase the burst length, but instead, we do not rearrange the data layout. The process of modifying the layout proposed in \cite{zhang2018caffeine} creates a certain complexity when writing a layer's output, given that the layer's output data has to be rearranged again to be accommodated to the next layer's input configuration (e.g., tile size). 

The work proposed by Alwani \textit{et al.} \cite{alwani2016fused} and Xiao \textit{et al.} \cite{xiao2017exploring} focuses on data-flow across multiple Conv-layers. Instead of processing a layer at a time, as usual, they focus on processing multiple layers at once without generating intermediate data between them. Their solution works by fusing multiple layers resulting in a computation pyramid across those layers. They use some complex data-structures to keep the intermediate data of each pyramid. In general, even reducing the memory transfers between the FPGA and host as they do, their accelerator still requires a huge amount of memory to store all the intermediate data from different pyramids. King \textit{et al.} \cite{kim2019simple}, on the other hand, proposes an algorithm to evaluate the scheduling of multiples Conv-layers from the start of a branch until the merge, thus keeping each layer's input/output data  as much as possible in on-chip memory. In \cite{kim2019simple}, they used an NPU with 1MB on-chip memory, which is enough for many cases, which is a bit expensive for edge inference AI accelerators.

It is also possible to reduce data transfers by applying data compression. NullHop \cite{aimar2018nullhop} does this in hardware, and Han \textit{et al.} \cite{han2015deep} does it by applying Huffman Coding. Sparsity is another technique used to avoid computing zero elements and therefore reducing data transfer of unnecessary data besides avoiding unnecessary computation. Such technique is used by several works \cite{han2015deep,li2019squeezeflow,han2016eie}. All these techniques could also be used to improve the approach proposed in this paper, although they are not the focus herein.

\section{Conclusion}\label{sec:conclusion}
Given the restricted on-chip memory sizes of NPU architectures, efficient data tiling and scheduling techniques are crucial to minimizing the cost of memory accesses. This paper proposes TSO, an optimization pass for the TF-XLA compiler that identifies the best combination of data tiling, scheduling and MAC operations that minimizes execution of convolutions in CNN models. To achieve that, TSO does a precise modeling of memory burst, achieving a speedup of up to 21.7\% for some typical CNN models when compared to no-burst modeling. TSO also achieves up to  41.0\% speedup when compared to a fixed TLE slicing, and 39.8\% when compared to a fixed TLT tiling. The TSO generality was also evaluated by porting and running it on the Glow toolchain.

\section*{Acknowledgment}
This work was supported by Institute of Information \& communications Technology Planning \& Evaluation (IITP) grants funded by the Korea government (MSIT) (No.2022-0-00769, Neuromorphic Computing Software Platform for Artificial Intelligence Systems), by Electronics and Telecommunications Research Institute (ETRI), and also by SilicoNeuro/AiM Future

\end{document}